\documentclass[aps,prd,twocolumn,nofootinbib]{revtex4-1}
\usepackage{epsfig}
\usepackage[colorlinks,linkcolor=blue,anchorcolor=blue,citecolor=blue,urlcolor=blue,breaklinks=true]{hyperref}
\usepackage{graphics}
\usepackage{slashed}
\usepackage{color}
\usepackage{amsmath}
\usepackage{bm}
\usepackage{setspace}
\usepackage{caption}
\usepackage{multirow}
\usepackage{subfigure}

\begin{document}
\author{Cheng-Ming Li$^{1,5}$}\email{licm.phys@gmail.com}
\author{Pei-Lin Yin$^{2}$}\email{yinpl@njupt.edu.cn}
\author{Hong-Shi Zong$^{1,3,4,5}$}\email{zonghs@nju.edu.cn}
\address{$^{1}$ Department of Physics, Nanjing University, Nanjing 210093, China}
\address{$^{2}$ College of Science, Nanjing University of Posts and Telecommunications, Nanjing, 210023, China}
\address{$^{3}$ Joint Center for Particle, Nuclear Physics and Cosmology, Nanjing 210093, China}
\address{$^{4}$ State Key Laboratory of Theoretical Physics, Institute of Theoretical Physics, CAS, Beijing, 100190, China}
\address{$^{5}$ Nanjing Proton Source Research and Design Center, Nanjing 210093, China}
\title{New algorithm to study the pseudo-Wigner solution of the quark gap equation in the framework of the (2+1)-flavor NJL model}

\begin{abstract}
In this paper, we study the pseudo-Wigner solution of the quark gap equation with a recently proposed algorithm in the framework of the (2+1)-flavor Nambu-Jona-Lasinio (NJL) model. We find that for the current quark mass $m_{\rm u,d}=5.5$ MeV and chemical potential $\mu<\mu_{\rm TCP}=272.5$ MeV, the Nambu solution and the positive pseudo-Wigner solution obtained via this algorithm is consistent with the physical solution obtained with the iterative method. Furthermore, the algorithm we used can help to illustrate the evolution of the solutions of the gap equation from the chiral limit to non-chiral limit and gives a prediction where the crossover line is located in the phase diagram for $\mu<272.5$ MeV. In addition, we also study the chiral susceptibilities as well as the loss of solutions for different chemical potentials.

\bigskip

\bigskip


\end{abstract}

% QED3 11.10.Kk, 11.15.Tk, 11.30.Qc
% quark-gluon, 12.38.Mh; Quark models, 12.39.-x; Quark confinement, 12.38.Aw; Quark deconfinement, 25.75.Nq; Lattice QCD calculations, 12.38.Gc;
% Quark-gluon plasma, 12.38.Mh; phase transitions in QGP, 25.75.Nq; production of QGP, 25.75.Nq; particle physics of chirality, 11.30.Rd;

\maketitle

\section{Introduction}
As a basic theory to describe the strong interacted matter and the dynamics of it, the quantum chromodynamics (QCD) plays an important role in the standard model of particle physics. For the low energy scale, the non-perturbative property is characterized in QCD, and the dynamical chiral symmetry breaking (DCSB) and color confinement are two critical phenomena in this scheme: the former explains the source of $98\%$ mass of the visible universe, and the latter is responsible for why the quark cannot be observed in the experiment. In fact, there are many approaches to study the QCD phase transition theoretically. As one of the most reliable one, the lattice QCD (LQCD)~\cite{Borsanyi2010,PhysRevLett.110.172001} confronts the "sign problem", making it difficult to carry out the calculations at finite chemical potential. Therefore, people recourse to the effective models such as the Dyson-Schwinger equations (DSEs)~\cite{ROBERTS1994477,Roberts2000S1,doi:10.1142/S0218301303001326,Cloet20141,PhysRevD.90.114031,PhysRevD.91.056003,PhysRevD.91.034017} and the Nambu-Jona-Lasinio (NJL) model~\cite{RevModPhys.64.649,Buballa2005205,Cui2013,NuclPhysB.896.682}. Some progresses have been made through these effective models in the study of the QCD phase transition and the QCD phase diagram~\cite{Roberts2000S1,PhysRevD.91.056003,PhysRevD.91.034017,Cui2013,NuclPhysB.896.682,PhysRevD.78.039902}.

As we know, the QCD phase diagram contains rich information: the upper left region of the diagram corresponds to the domain of the thermal QCD where the temperature is very high, and the features of the early universe as well as its expansion can be addressed here~\cite{PhysRevLett.105.041301,ENQVIST1993298}; the lower right region of the diagram corresponds to the domain of the dense QCD where the chemical potential is very large, and the study of the neutron star is implemented here~\cite{2018ApJ...860...57A,PhysRevD.98.083013,PhysRevD.97.103013,PhysRevD.95.056018,PhysRevD.86.114028,PhysRevC.83.025805,doi:10.1142/S0218301313500262}. At present, a popular scenario is in favor of the existence of the critical end point (CEP)~\cite{doi:10.1142/S0217751X17500671,10.1038/srep45937,PhysRevD.93.114014,PhysRevD.93.034013,AYALA201577,KOHYAMA2015682}: for the chemical potential smaller (larger) than this point, as the temperature increases, the QCD system will confront a crossover (first order phase transition). Analogously, in the case of chiral limit where the current quark mass $m=0$, the tri-critical point (TCP) is favored: for the chemical potential smaller (larger) than this point, as the temperature increases, the QCD system will experience a second order phase transition (first order phase transition). However, whether the CEP exists and where it is located are still open questions and model dependent. In addition, the relation between the deconfinement phase transition and the chiral phase transition is also unknown currently, but people believe that they are tightly connected~\cite{Roberts2000S1,MARQUEZ2015529}. For many studies~\cite{Cui2018,PhysRevD.76.074023,0004-637X-836-1-89}, these two kinds of transitions are regarded to happen simultaneously and the rigorous distinction between them is ignored. In this work, we will also employ this viewpoint.

It is well known that the gap equation can be strictly solved in the framework of the (2+1)-flavor NJL model with many methods, such as the graphical and the iterative method. And in the case of the chiral limit, there are two important solutions to the quark's gap equation, namely the Nambu solution and the Wigner solution. The physical meaning of these two solutions is very clear, i.e., the stable state with the breaking or restoring of the chiral symmetry. However, by solving the quark's gap equation in the case of non-chiral limit, we have found multiple solutions, which has been studied by a variety of studies within the two-flavor effective models~\cite{0256-307X-22-12-014,PhysRevC.75.015201,PhysRevD.86.114001,WILLIAMS2007167,Cui2013,Cui2014,PhysRevD.88.096003,PhysRevD.94.096003,PhysRevC.63.025202,0256-307X-29-4-041201,PhysRevLett.106.172301,Cui2018,0954-3899-45-10-105001}. In addition to the physical meaning of the Nambu solution, the physical meaning of other solutions is unclear (especially for the solution which is closely related to the Wigner solution in the case of chiral limit), because the chiral symmetry is explicitly broken and only can be partially restored in the non-chiral limit. We believe that if we do not solve this problem, it is meaningless to discuss the chiral phase transition of strong interacted matter in an effective model, such as the NJL model in this work. On the other hand, in the crossover region of the QCD diagram (for $T>T_E=48$ MeV or $\mu<\mu_E=324$ MeV in the (2+1)-flavor NJL model with the Hatsuda-Kunihiro parameters~\cite{HATSUDA1994221,PhysRevD.77.114028}), the physical solution of the quark gap equation is positive and decreases smoothly to the nonzero current quark mass as the temperature or chemical potential increases to infinity. Then a question naturally arises: When the pseudo-Wigner solution (see the description at the end of Sec.~\ref{one} for the definition of the pseudo-Wigner solution) emerges? In the past, people used to utilize the QCD susceptibilities (the linear responses of the quark condensate to the external fields) to give a criterion~\cite{PhysRevD.88.114019,Lu2015,CUI2015172}, but the pseudo-Wigner solution in the non-chiral case is still not clarified.

In a word, only if we find out the pseudo-Wigner solution of the gap eqution in the non-chiral limit and figure out its relation to the Wigner solution in the chiral limit, the statement can make sense that the phase of strong interacting matter evolves from Nambu phase to pseudo-Wigner phase as the temperature and chemical potential increase. The main purpose of this paper is to study how the Wigner solution of the quark gap equation evolves with the current quark mass in the case of chiral limit to the so-called pseudo-Wigner solution in the non-chiral limit with the (2+1)-flavor NJL model. To achieve the above objectives, we only use the algebraic method provided by the Ref.~\cite{0954-3899-45-10-105001}, because this algebraic method can express the Wigner solutions evolution with the current quark mass from the case of the chiral limit to non-chiral limit clearly and directly.

It is noted that the previous studies also throw light on the loss of solutions of the gap equation~\cite{PhysRevD.92.125035,PhysRevC.75.015201}. Similar to the analysis in these papers, the algorithm we employ in this paper also meet this issue when dealing with the evolution from the Wigner solution to the pseudo-Wigner solution, because the derivatives of the effective quark masses $M_u', M_s'$ might diverge for some values of $(T, \mu)$ as the current quark mass increases. As a result, the Nambu and pseudo-Wigner solution evolved from the chiral limit might fail to cover the strict physical solution in non-chiral limit, thus implying a region of the QCD phase diagram inaccessible with this algorithm.

This paper is organized as follows: In Sec.~\ref{one}, we briefly discuss some contemporary perspectives of the dynamical chiral symmetry breaking and restoring within the (2+1)-flavor NJL model. In Sec.~\ref{two}, the recently proposed algorithm~\cite{0954-3899-45-10-105001} is applied in the study of the solutions (especially the pseudo-Wigner solution) of the quark gap equation with nonzero current quark mass for finite temperature and chemical potential. Then we present the loss of solutions in different schemes. For a better understanding of the evolution of the solution, the chiral susceptibilities as well as a prediction of part of the QCD phase diagram with our algorithm are also calculated in this section. Finally, a brief summary and discussion are given in Sec.~\ref{three}.

\section{Dynamical breaking and restoring of the chiral symmetry within 2+1 flavors NJL model}\label{one}
As one of the most typical phenomena in the non-perturbative QCD, chiral symmetry breaking and restoring is a hot issue at present, and the study of it helps a lot for the developing of the quark model. Compared with the basic energy scale $\Lambda_{\rm QCD}$, the current $u (d)$-quark mass is very small, which is estimated to be $~3.5^{+0.5}_{-0.2}$ MeV in Ref.~\cite{PhysRevD.98.030001}, thus the QCD Lagrangian shows an approximate chiral symmetry for the flavor SU(2) version. For the flavor SU(3) version, the current $s$-quark mass is much larger than that of $u (d)$-quark ($~95^{+9}_{-3}$ MeV in Ref.~\cite{PhysRevD.98.030001}), then its contribution to the chiral symmetry breaking cannot be neglected. Actually, if the current quark mass is nonzero, no matter how small it is, the chiral symmetry is explicitly broken. Oppositely, in the case of chiral limit where the current $u (d)$-quark mass is fixed to be zero, the chiral symmetry is preserved in the Lagrangian. But the chiral symmetry of the QCD system is still broken in this case due to the dynamical (or "spontaneously" for some different profiles) chiral symmetry breaking mechanism. It is noted that this mechanism is different from the Higgs mechanism, because the chiral symmetry is a global symmetry whose current does not couple to the gauge fields. Besides, the QCD vacua are distinguishable with the order parameter's different orientations.

In the past, the QCD vacuum is considered as a condensed state of quark-antiquark pairs, and its corresponding vacuum expectation value $\langle\bar{\psi}\psi\rangle=\bar{\psi}_{\rm R}\psi_{\rm L}+\bar{\psi}_{\rm L}\psi_{\rm R}$ (also denoted as $\phi$ sometimes) is regarded as a good indicator of the chiral symmetry. However, there is a popular perspective that the quark condensate is actually a property of hadrons~\cite{PhysRevC.82.022201,PhysRevC.85.065202}, thus strongly interacted with the quarks confined in hadrons. From this viewpoint, the chiral symmetry and its breaking can be reflected by the constituent (effective) quark masses. Nevertheless, in the regime of strong coupling strength, the perturbative QCD is invalid and non-perturbative approaches are needed. In this work, we will employ the (2+1)-flavor NJL model to study the chiral phase transition via the solutions of the quark gap equation.

The NJL model is a good candidate for the study of QCD with low energy scales. Using the same notations as Ref.~\cite{HATSUDA1994221}, its Lagrangian of 2+1 flavors reads\footnote{We will work in the Euclidean Space throughout this paper, and the exact isospin symmetry between the $u$ and $d$ quark is utilized, that is, $m_{\rm u}=m_{\rm d}\equiv m$. And from the following calculations in this paper we can see that $M_{\rm u}=M_{\rm d}$, $\phi_{\rm u}=\phi_{\rm d}$.}:
\begin{equation}\label{NJLlagrangian}
\mathcal{L}_{\rm NJL}=\bar{\psi}(i{\not\!\partial}-m)\psi+\mathcal{L}_{S}+\mathcal{L}_{A},
\end{equation}
with
\begin{equation}\label{fourpointinteraction}
\mathcal{L}_{S}=\frac{g_{\rm S}}{2}[(\bar{\psi}\lambda_a\psi)^2+(\bar{\psi}{\rm i}\gamma_5\lambda_a\psi)^2],
\end{equation}
and
\begin{equation}\label{tHooftterm}
\mathcal{L}_{A}=g_{\rm D}[{\rm det}\bar{\psi}(1-\gamma_5)\psi+{\rm H.c.}].
\end{equation}
Here $\lambda_a$ is the Gell-Mann matrix in flavor space and $\lambda_0=\sqrt{\frac{2}{3}}I$. $m={\rm diag}(m_{\rm u},m_{\rm d},m_{\rm s})$ represents the matrix of the current quark mass. Under the mean-field approximation, the gap equations can be written as,
\begin{equation}\label{gapeq}
  M_{\rm i}=m_{\rm i}-2g_{\rm S}\phi_{\rm i}-2g_{\rm D}\phi_{\rm j}\phi_{\rm k},
\end{equation}
where $\phi_{\rm i}$ is the quark condensate of flavor i and $({\rm i,\,j,\,k})=$any permutation of $(u,\,d,\,s)$. To proceed the following calculations, a certain regularization should be employed. In this work, we will use the three-momentum cutoff regularization with an ultraviolet cutoff $\Lambda$. By definition, the quark condensate can be expressed as,
\begin{eqnarray}
 \phi_{\rm i} =&& -\int\frac{{\rm d}^4p}{(2\pi)^4}{\rm Tr}[S^{\rm i}(p)]\nonumber\\
  =&&- N_{\rm c}\int_{-\infty}^{+\infty}\frac{{\rm d}^4p}{(2\pi)^4}\frac{4M_{\rm i}}{p^2+M_{\rm i}^2}\nonumber\\
  =&&-\frac{N_{\rm c}M_{\rm i}}{\pi^2}\int_{0}^{\Lambda}{\rm d}pp^2\big[1-(e^{\frac{E_{p,{\rm i}}-\mu}{T}}+1)^{-1}\nonumber\\
  &&-(e^{\frac{E_{p,{\rm i}}+\mu}{T}}+1)^{-1}\big]/E_{p,{\rm i}},\,\,\label{qcondensate}
\end{eqnarray}
where the trace "Tr" is taken in the Dirac and color spaces and $N_{\rm c}=3$ is the number of colors. $S^{\rm i}(p)=\frac{1}{{\rm i}{\not p}+M_{\rm i}}$ and $E_{p,{\rm i}}=\sqrt{\overrightarrow{p}^2+M_{\rm i}^2}$ represents the dressed quark propagator and the on-shell quark energy of flavor i, respectively. Then the thermodynamic potential (which is also called the total free energy) has the following form,
\begin{eqnarray}
% \nonumber % Remove numbering (before each equation)
  \Omega = &&g_{\rm S}(\phi_{\rm u}^2+\phi_{\rm d}^2+\phi_{\rm s}^2)+4g_{\rm D}\phi_{\rm u}\phi_{\rm d}\phi_{\rm s}\nonumber\\
  &&-2N_{\rm c}\sum_{\rm i=u,d,s}\int_{0}^{\Lambda}\frac{{\rm d}^3p}{(2\pi)^3}\bigg\{E_{p,{\rm i}}+T{\rm ln}\bigg[1+e^{\frac{-(E_{p,{\rm i}}-\mu)}{T}}\bigg]\nonumber\\
  &&+T{\rm ln}\bigg[1+e^{\frac{-(E_{p,{\rm i}}+\mu)}{T}}\bigg]\bigg\},\label{thermdynpotential}
\end{eqnarray}
In general, the quark gap Eq.~(\ref{gapeq}) can be solved with many methods, such as the iterative method and the graphical method. Also, the solutions can be obtained via searching for the extreme value of the thermodynamic potential. In this way, we can see not only the quantities of the solutions but also the positions of them more directly. In this work, the parameter set employed is shown in Table.~\ref{parameters}, same as that in Ref.~\cite{HATSUDA1994221} for the case of 2+1 flavors.
\begin{table}
\caption{Parameter set fixed in our work.}\label{parameters}
\begin{tabular}{p{1.1cm}p{1.1cm}p{1.1cm}p{2.2cm}p{2.2cm}}
%\begin{tabular}{cccccc}
\hline\hline
$m$&$m_{\rm s}$&$\Lambda$&$g_{\rm S}$&$g_{\rm D}$\\
$[{\rm MeV}]$&$[{\rm MeV}]$&$[{\rm MeV}]$&$[{\rm MeV}^{-2}]$&$[{\rm MeV}^{-5}]$\\
\hline
5.5&135.7&631.4&$9.21\times10^{-6}$&$-9.26\times10^{-14}$\\
\hline\hline
\end{tabular}
\end{table}

\begin{figure}
\includegraphics[width=0.47\textwidth]{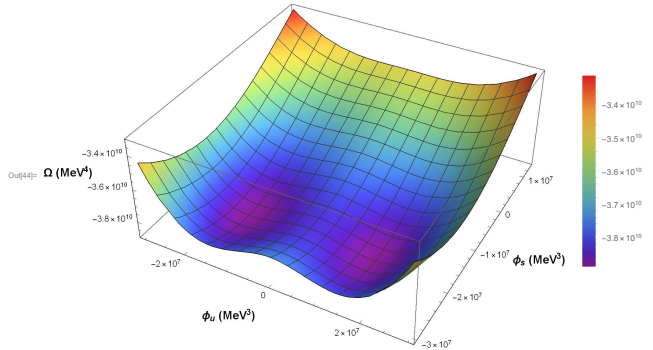}
\caption{The thermodynamic potential in the chiral limit for $T=0$ and $\mu=0$.}
\label{Fig:T0mu0m0}
\end{figure}
\begin{figure}
\includegraphics[width=0.47\textwidth]{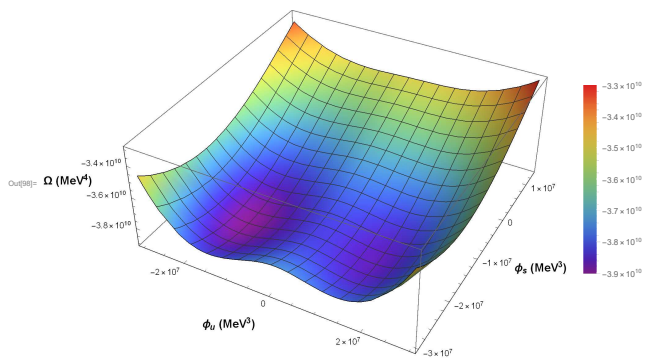}
\caption{The thermodynamic potential in the case of current quark mass $m=5.5$ MeV for $T=0$ and $\mu=0$.}
\label{Fig:T0mu0m55}
\end{figure}
In many cases, the gap Eq.~(\ref{gapeq}) has more than one solution, meaning that we will encounter the issue of multiple solutions. For example, in Fig.~\ref{Fig:T0mu0m0} and Fig.~\ref{Fig:T0mu0m55}, we can see that for zero temperature and chemical potential, no matter in the chiral limit case or in the case of $m=5.5$ MeV, there are three extreme values of $\Omega$: two valley points and one saddle point, corresponding to three numerical solutions of the gap Eq.~(\ref{gapeq}) in this scheme. Just as the declaration in Refs.~\cite{PhysRevC.75.015201,PhysRevD.86.114001,Cui2018,0954-3899-45-10-105001}, the negative solution of $M_{\rm u,d}$ is only a mathematical solution with no physical meanings. Therefore we will only study the solutions with $M_{\rm u,d}\geq0$.

In the chiral limit, the Nambu solution ($M_{\rm u,d}>0$ with a local minimum thermodynamic potential) represents the breaking of the chiral symmetry, and can be distinguished from the pseudo-Nambu solution ($M_{\rm u,d}>0$ with a local maximum thermodynamic potential) and the Wigner solution ($M_{\rm u,d}=0$, thus the chiral symmetry is restored). But in the non-chiral limit, things become complicated because the chiral symmetry is explicitly broken and multiple solutions are always found. For low temperature and high chemical potential where the first-order chiral phase transition happens, the solution with a biggest $M_{\rm u,d}$ as well as a local minimum thermodynamic potential is still called the Nambu solution, while the solution with a smallest $M_{\rm u,d}$ and a local minimum thermodynamic potential is always called the pseudo-Wigner solution where the chiral symmetry is partially restored\footnote{In principle, the pseudo-Wigner solution should also be physical with $M_{\rm u,d}\geq0$. However, to clarify the whole evolution process from the Wigner solution in the chiral limit to the pseudo-Wigner solution in non-chiral limit with our algorithm in this paper, both the physical and unphysical pseudo-Wigner solutions are collectively referred to as the pseudo-Wigner solution.}. Then how about the scheme with high temperature and low chemical potential where no chiral phase transition but a crossover happens? As we know, there is only one solution satisfying $M_{\rm u,d}>0$ in this case, thus it is hard to figure out when the chiral symmetry is partially restored directly. In the past, people used to resort to various linear responses of the quark condensate, i.e., the susceptibilities to determine the starting point of the pseudo-Wigner solution in this case.

In the following of this work, we will employ a new algorithm to solve this problem. Furthermore, the solutions of the gap equation in the non-chiral limit can be related naturally to that in the chiral limit with this algorithm. And the loss of solutions as well as the QCD phase diagram are also discussed in this work.

\section{Solutions of the quark gap equation with the new algorithm}\label{two}
In this section, we will introduce a recently proposed algorithm to explore the solutions of the quark gap equation in the non chiral limit case, especially the pseudo-Wigner solution which is difficult to ascertain at low chemical potential and high temperature. In fact, for a certain temperature and chemical potential, we can regard the effective quark masses $M_{\rm u}$ and $M_{\rm s}$ as functions of the current $u (d)$-quark mass $m$, and the derivative of Eq.~(\ref{gapeq}) can be derived as,
\begin{eqnarray}
% \nonumber % Remove numbering (before each equation)
  M_{\rm u}' &=& 1-2g_{\rm S}\phi_{\rm u}'-2g_{\rm D}(\phi_{\rm u}'\phi_{\rm s}+\phi_{\rm u}\phi_{\rm s}'),\,\,\label{uderivgeq}\\
  M_{\rm s}' &=& -2g_{\rm S}\phi_{\rm s}'-4g_{\rm D}\phi_{\rm u}'\phi_{\rm u}.\,\,\label{sderivgeq}
\end{eqnarray}
The derivative of the quark condensate is
\begin{eqnarray}
% \nonumber % Remove numbering (before each equation)
  \phi_{\rm i}' =&& -\frac{N_{\rm c}M_{\rm i}'}{\pi^2}\int_{0}^{\Lambda}{\rm d}p\bigg\{\frac{p^4}{(p^2+M_{\rm i}^2)^{\frac{3}{2}}}\nonumber\\
  &&+\frac{M_{\rm i}^2p^2}{T(p^2+M_{\rm i}^2)}\bigg[\frac{e^{\frac{E_{p,{\rm i}}-\mu}{T}}}{(e^{\frac{E_{p,{\rm i}}-\mu}{T}}+1)^2}+\frac{e^{\frac{E_{p,{\rm i}}+\mu}{T}}}{(e^{\frac{E_{p,{\rm i}}+\mu}{T}}+1)^2}\bigg]\nonumber\\
  &&-\frac{p^4}{(p^2+M_{\rm i}^2)^{\frac{3}{2}}}\bigg[\frac{1}{e^{\frac{E_{p,{\rm i}}-\mu}{T}}+1}+\frac{1}{e^{\frac{E_{p,{\rm i}}+\mu}{T}}+1}\bigg]\bigg\}.\,\,\label{condensatederiv}
\end{eqnarray}
We can find that the derivative of the quark condensate is proportional to the derivative of the effective quark mass. Thus Eq.~(\ref{uderivgeq}) and Eq.~(\ref{sderivgeq}) are actually coupled linear equations of $M_{\rm u}'$ and $M_{\rm s}'$. Assuming we have already known the Wigner or pseudo-Wigner solution of Eq.~(\ref{gapeq}), then we can substitute it into Eq.~(\ref{uderivgeq}) and Eq.~(\ref{sderivgeq}) to get $M_{\rm u}'$ and $M_{\rm s}'$. If the matrix of $\frac{\delta\{M_{\rm u}',M_{\rm s}'\}}{\delta\{M_{\rm u},M_{\rm s}\}}$ is not singular, the coupled Eq.~(\ref{uderivgeq}) and Eq.~(\ref{sderivgeq}) will only have one solution. In other words, if the matrix above is singular for a certain ($\mu, T$), the derivatives $M_{\rm u}'$ and $M_{\rm s}'$ will become infinity, resulting in a invalidation of our algorithm under this scheme.

With the Wigner solution of Eq.~(\ref{gapeq}) in the chiral limit (which is actually the trivial solution by the analysis in Sec.~\ref{one}) and its corresponding derivative, the pseudo-Wigner solution of the gap equation for non chiral limit can be derived via the following equations,
\begin{eqnarray}
% \nonumber % Remove numbering (before each equation)
  M_{\rm u}^W(m) &=& M_{\rm u}^W(0)+\int_{0}^{m}{\rm d}\tilde{m}M_{\rm u}'(\tilde{m}),\,\,\label{Mufornonchirlim}\\
  M_{\rm s}^W(m) &=& M_{\rm s}^W(0)+\int_{0}^{m}{\rm d}\tilde{m}M_{\rm s}'(\tilde{m}).\,\,\label{Msfornonchirlim}
\end{eqnarray}
From Eqs.~(\ref{Mufornonchirlim}) and (\ref{Msfornonchirlim}), we can see that this algorithm also requires the continuity of $M_{\rm u}'$ and $M_{\rm s}'$ with the increasing of $m$. If we discretize the range of $\tilde{m}$ to $N$ grids, the integral will approximate to the summation with an interval of $\Delta m=m/N$. In this work, we set the number of discretization points $N=550$. Then the pseudo-Wigner solution for small $\Delta m=0.01$ MeV is
\begin{eqnarray}
% \nonumber % Remove numbering (before each equation)
  M_{\rm u}^W(\Delta m) &=& M_{\rm u}^W(0)+M_{\rm u}^{'W}(0)\Delta m = M_{\rm u}^{'W}(0)\Delta m,\,\,\,\,\,\,\label{Muclforsmallm}\\
  M_{\rm s}^W(\Delta m) &=& M_{\rm s}^W(0)+M_{\rm s}^{'W}(0)\Delta m.\,\,\label{Msclforsmallm}
\end{eqnarray}
For the mass $m=n\Delta m, n\geq1$,
\begin{eqnarray}
% \nonumber % Remove numbering (before each equation)
  M_{\rm u}^W(n\Delta m) &=& M_{\rm u}^W((n-1)\Delta m)+M_{\rm u}^{'W}((n-1)\Delta m)\Delta m,\,\,\,\,\,\,\,\,\,\,\label{Muclforndeltam}\\
  M_{\rm s}^W(n\Delta m) &=& M_{\rm s}^W((n-1)\Delta m)+M_{\rm s}^{'W}((n-1)\Delta m)\Delta m.\,\,\,\,\,\,\,\,\,\,\label{Msclforndeltam}
\end{eqnarray}
\begin{figure}
\includegraphics[width=0.47\textwidth]{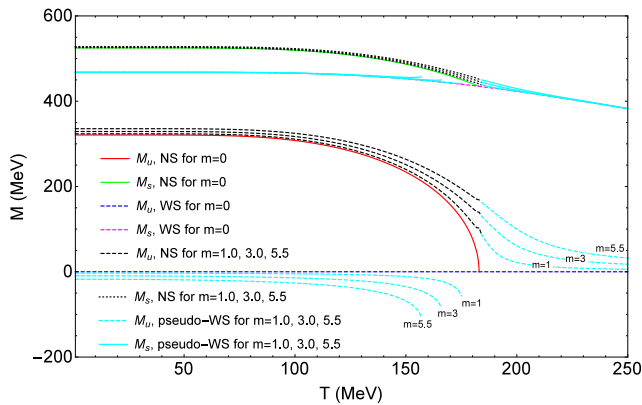}
\caption{The evolution of solutions of the quark gap equation with our algorithm from the chiral limit to non-chiral limit for $\mu=0$ and $T\neq0$. The effective quark mass $M_u$ and $M_s$ of the Nambu solution in the chiral limit is shown in the red solid line and green solid line, respectively. The $M_u$ and $M_s$ of the Wigner solution in the chiral limit is shown in the blue dashed line and magenta dashed line, respectively. The $M_u$ and $M_s$ of the Nambu solution for the current $u, d$ quark mass $m=1.0, 3.0, 5.5$ MeV is shown in the black dashed line and black dotted line, respectively. The $M_u$ and $M_s$ of the pseudo-Wigner solution for the current $u, d$ quark mass $m=1.0, 3.0, 5.5$ MeV is shown in the cyan dashed line and cyan solid line, respectively.}
\label{Fig:TMuMsevolutionmu0}
\end{figure}
%\begin{figure}
%\includegraphics[width=0.47\textwidth]{thermpotmu0.eps}
%\caption{The thermodynamic potentials of the Nambu and Wigner solution in the chiral limit for $\mu=0$ and $T\neq0$, shown in the red solid line and red dotted line, respectively.}
%\label{Fig:thermpotmu0}
%\end{figure}

In addition, the algorithm above can also be applied to finding the Nambu solution of the quark gap equation in the case of non chiral limit, which can test the validity of the algorithm from the other side.
In Fig.~\ref{Fig:TMuMsevolutionmu0}, we study the evolution of solutions of the quark gap equation from the chiral limit to non-chiral limit for $\mu=0, T\neq0$. In this figure, we can see that there is a second-order chiral phase transition at $T=183.3$ MeV in the chiral limit, and for $T>183.3$ MeV, the Nambu solution vanishes. As the current quark mass $m$ increases, the pseudo-Wigner solution for $T<183.3$ MeV is going to be negative, but for $T>183.3$ MeV, it turns to be positive and connects well to the Nambu solution. As a result, the chiral phase transition is second-order in the chiral limit, but a crossover in non-chiral limit.

\begin{figure}
\includegraphics[width=0.47\textwidth]{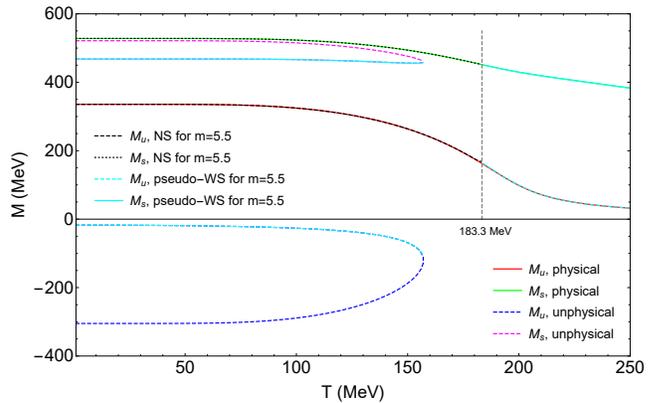}
\caption{The comparison of the solutions of the gap equation with our algorithm and with the iterative method in the scheme of $m=5.5$ MeV for $\mu=0$ and $T\neq0$. The strict physical solution of $M_u$ and $M_s$ is shown in the red solid line and green solid line, respectively. The strict unphysical solution of $M_u$ and $M_s$ is shown in the blue dashed line and magenta dashed line, respectively. The other lines in this figure is shown in the same plot type as in Fig.~\ref{Fig:TMuMsevolutionmu0}.}
\label{Fig:comparisonformu0}
\end{figure}
\begin{figure}
\includegraphics[width=0.47\textwidth]{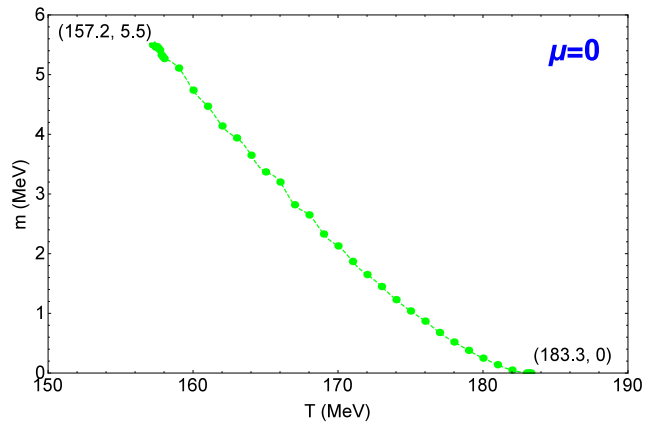}
\caption{The loss of pseudo-Wigner solution in our algorithm, shown in the $m-T$ plane for $\mu=0$. The green line means that in the scheme of $\mu=0$, for a certain ($T, m$), the derivative $M_{\rm u}'$ is going to be infinity, thus the algorithm is invalid to give the pseudo-Wigner solution above the green line in this diagram.}
\label{Fig:lossofsolutionsmu0}
\end{figure}
In Fig.~\ref{Fig:comparisonformu0}, we compare the strict solution of the gap equation with our result in the non-chiral limit. We can see that for $\mu=0$ and $m=5.5$ MeV, the Nambu and positive pseudo-Wigner solution obtained with our algorithm are in good agreement with the physical solution obtained with the iterative method. And to clarify the loss of solutions in the evolution from the chiral limit to non-chiral limit, we draw Fig.~\ref{Fig:lossofsolutionsmu0}. In this figure, the loss of pseudo-Wigner solution is shown in the $m-T$ plane for $\mu=0$. We present the diverge points of the derivative $M_{\rm u}'$ in the green line. Above this line, our algorithm is invalid to give the pseudo-Wigner solution. From this figure, we can see that in the scheme of $\mu=0$, the algorithm we employed will encounter the loss of pseudo-Wigner solution for $T\in(157.4,183.3)$ MeV through the evolution from the chiral limit to $m=5.5$ MeV.

\begin{figure}
\includegraphics[width=0.47\textwidth]{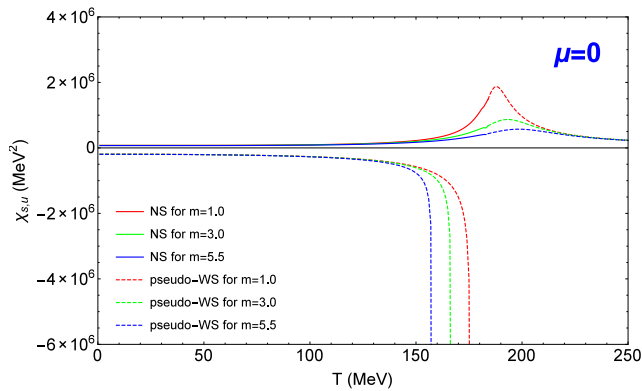}
\caption{The chiral susceptibilities of $u, d$ quark in the non-chiral limit cases obtained with our algorithm for $\mu=0$ and $T\neq0$. The red solid line, green solid line and blue solid line corresponds to the chiral susceptibility of the Nambu solution for $m=1.0, 3.0, 5.5$ MeV, respectively, while the dashed line in the same color represents the chiral susceptibility of the pseudo-Wigner solution in the same case.}
\label{Fig:chiralsusofumu0}
\end{figure}
Experimentally, for a certain physical system, its linear responses to the external field are always measured to investigate the properties of the system, such as the conductivity and susceptibility. Therefore, besides the calculation of the effective quark masses, we will also calculate the chiral susceptibilities for a better understanding of the solution of the quark gap equation. According to the definition, the chiral susceptibility of flavor i has the following form,
\begin{equation}\label{chiralsus}
  \chi_{s,{\rm i}} = -\frac{\partial\phi_{\rm i}}{\partial m}.
\end{equation}
We present the chiral susceptibilities of $u, d$ quark for $\mu=0$ in Fig.~\ref{Fig:chiralsusofumu0}. From this figure, we can find that in different schemes, the chiral susceptibilities of the Nambu and positive pseudo-Wigner solution keep finite, but the negative pseudo-Wigner solutions diverge at different temperatures. It is noted that the diverging points in this diagram can also be found in the green line of Fig.~\ref{Fig:lossofsolutionsmu0}, i.e., the points ($T, m$) = (176, 1.0), (167, 3.0), (157.2, 5.5) MeV on the $m-T$ plane. Actually, from Eq.~(\ref{condensatederiv}) we can see that if the derivative $M_{\rm u}'$ diverges, $\phi_{\rm u}'$ will also diverge, which is just the opposite number of the chiral susceptibility of $u, d$ quark.

Then we extend our calculation to the finite chemical potential. In fact, if the chemical potential is not so large that the chiral phase transition is still a second-order phase transition in the chiral limit, the evolution of the solutions of the gap equation will be different only quantitatively, but qualitatively unchanged. However, if the chemical potential is large enough that $\mu>\mu_{\rm TCP}=272.5$ MeV, things will become thoroughly different. Specifically, we will set $\mu=300$ MeV in the following to study the evolution of the solutions of the gap equation with our algorithm.

\begin{figure}
\includegraphics[width=0.47\textwidth]{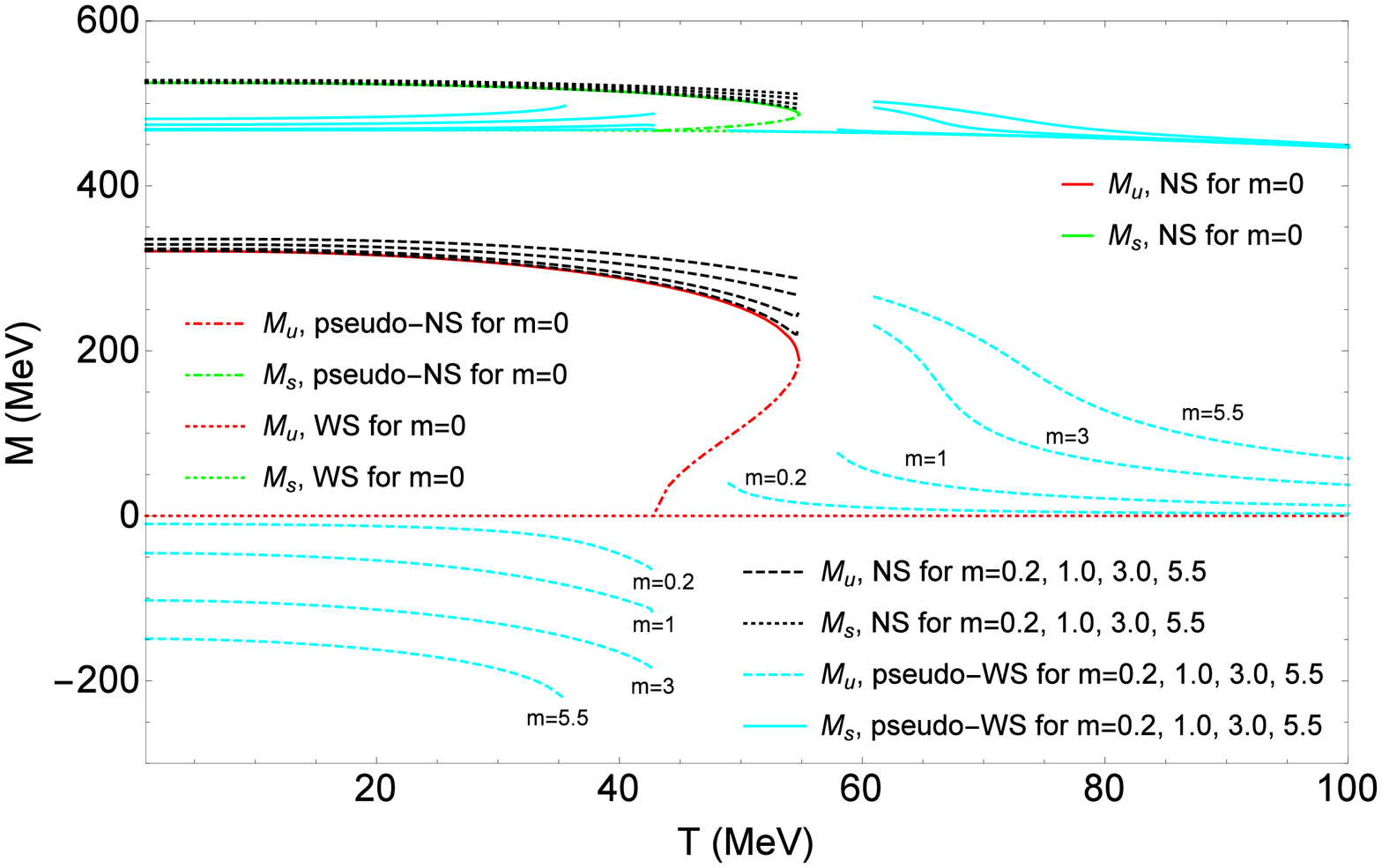}
\caption{The evolution of solutions of the quark gap equation with our algorithm from the chiral limit to non-chiral limit for $\mu=300$ MeV and $T\neq0$. The effective quark mass $M_u$ and $M_s$ of the pseudo-Nambu solution in the chiral limit is shown in the red dotdashed line and green dotdashed line, respectively. The $M_u$ and $M_s$ of the Wigner solution in the chiral limit is shown in the red dotted line and green dotted line, respectively. The other lines in this figure is shown in the same plot type as in Fig.~\ref{Fig:TMuMsevolutionmu0}.}
\label{Fig:TMuMsevolutionmu300}
\end{figure}
\begin{figure}
\includegraphics[width=0.47\textwidth]{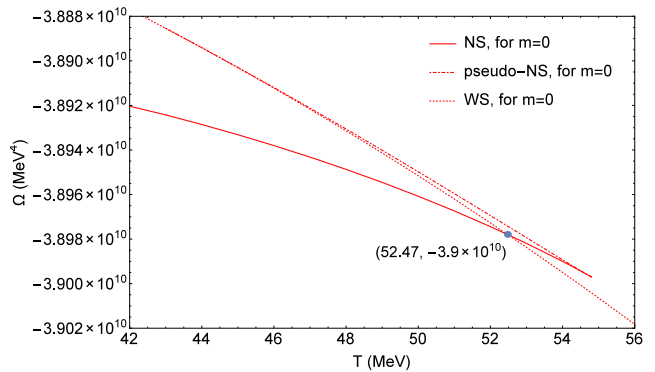}
\caption{The thermodynamic potentials of the Nambu, pseudo-Nambu and Wigner solution in the chiral limit for $\mu=300$ MeV and $T\neq0$, shown in the red solid line, red dotdashed line and red dotted line, respectively.}
\label{Fig:thermpotmu300}
\end{figure}
As we know, for $\mu>\mu_{\rm TCP}$, the QCD system in the chiral limit will encounter a first-order chiral phase transition as the temperature increases. In the region of multiple solutions, the Nambu, pseudo-Nambu, and Wigner solutions coexist, just as the red solid line, red dotdashed line, and the red dotted line shown in Fig.~\ref{Fig:TMuMsevolutionmu300}. However, the pseudo-Nambu solution has a local maximum thermodynamic potential, thus does not corresponds to a physically stable state. In Fig.~\ref{Fig:thermpotmu300} we can see that the potential of the pseudo-Nambu solution is the highest of three potentials, therefore we will not consider the evolution process of the pseudo-Nambu solution.

In Fig.~\ref{Fig:TMuMsevolutionmu300}, the Wigner solution and pseudo-Nambu solution converge at about $T=42.8$ MeV in the chiral limit, and the pseudo-Wigner solution evolves to two branches separated by this point in the non-chiral limit. As $m$ increases, the distance of the positive and negative pseudo-Wigner solution is going to be farther and farther. Although the Nambu solution does not encounter the loss of solution in the evolution from the chiral limit to non-chiral limit, it just cannot connect to the positive pseudo-Wigner solution for any $m$ like that in Fig.~\ref{Fig:TMuMsevolutionmu0}.

\begin{figure}
\includegraphics[width=0.47\textwidth]{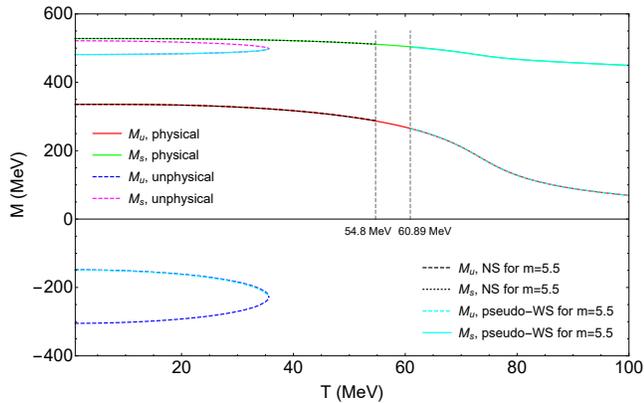}
\caption{The comparison of the solutions of the gap equation with our algorithm and with the iterative method in the scheme of $m=5.5$ MeV for $\mu=300$ MeV and $T\neq0$. The lines in this figure is shown in the same plot type as in Fig.~\ref{Fig:comparisonformu0}.}
\label{Fig:comparisonformu300}
\end{figure}
In Fig.~\ref{Fig:comparisonformu300} we compare the solutions of the gap equation with our algorithm and with the iterative method for $m=5.5$ MeV. And the solution is found to be absent in the region of $T\in(54.8, 60.89)$ MeV with our algorithm, thus cannot cover the strict physical solution and results in the invalidation of our algorithm in this scheme.

\begin{figure}
\includegraphics[width=0.47\textwidth]{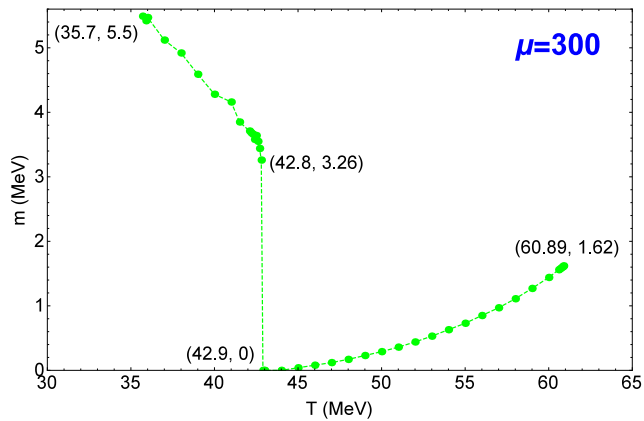}
\caption{The loss of pseudo-Wigner solution in our algorithm, shown in the $m-T$ plane for $\mu=300$ MeV. The green line in this figure has the same meaning as in Fig.~\ref{Fig:lossofsolutionsmu0}.}
\label{Fig:lossofsolutionsmu300}
\end{figure}
To check the loss of solutions in the evolution process, we plot Fig.~\ref{Fig:lossofsolutionsmu300}, and the green line in this figure represents the same meaning as in Fig.~\ref{Fig:lossofsolutionsmu0}. The difference is that there is a sharp gap in the green line for $\mu=300$ MeV, demonstrating a different process of the loss of solutions as $m$ increases. For $m<1.62$ MeV, the positive pseudo-Wigner solution will experience the loss of solution as $m$ increases; for $1.62<m<3.26$ MeV, the loss region of the solution keeps almost unchanged; and for $3.26<m<5.5$ MeV, the negative pseudo-Wigner solution starts to undergo the loss of solutions while the loss region of the positive pseudo-Wigner solution does not change in this case. During the evolution from the chiral limit to $m=5.5$ MeV, the pseudo-Wigner solution will encounter a total loss range of $T\in(35.7, 60.89)$ MeV. And we can imagine that for other chemical potentials larger than 272.5 MeV, similar things will happen as that for $\mu=300$ MeV.

\begin{figure}
\includegraphics[width=0.47\textwidth]{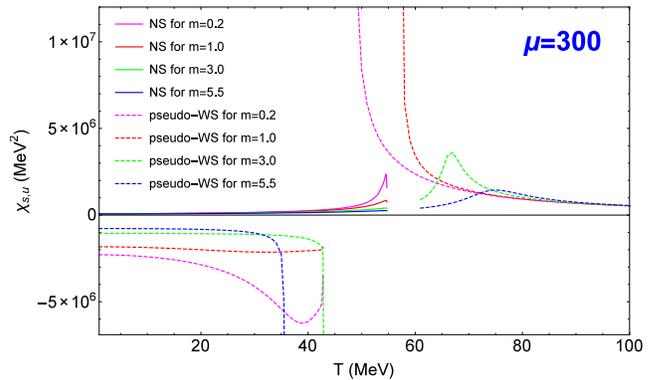}
\caption{The chiral susceptibilities of $u, d$ quark in the non-chiral limit cases obtained with our algorithm for $\mu=300$ MeV and $T\neq0$. The magenta solid line, red solid line, green solid line and blue solid line corresponds to the chiral susceptibility of the Nambu solution for $m=0.2, 1.0, 3.0, 5.5$ MeV, respectively, while the dashed line in the same color represents the chiral susceptibility of the pseudo-Wigner solution in the same case.}
\label{Fig:chiralsusofumu300}
\end{figure}
On the other hand, we also present the chiral susceptibilities of $u, d$ quark to study the loss of solutions for $\mu=300$ MeV, which can be found in Fig.~\ref{Fig:chiralsusofumu300}. In this figure, when $m$ is small, the $\chi_{s,u}$ of the positive pseudo-Wigner solution diverges at some temperature but the negative branch does not. As $m$ increases to a relatively large amount, the positive branch finally keeps finite but the negative branch begins to diverge. Unlike the case of $\mu=0$, during the whole evolution process, the $\chi_{s,u}$ of the positive pseudo-Wigner solution does not connect to that of the Nambu solution.

\begin{figure}
\includegraphics[width=0.47\textwidth]{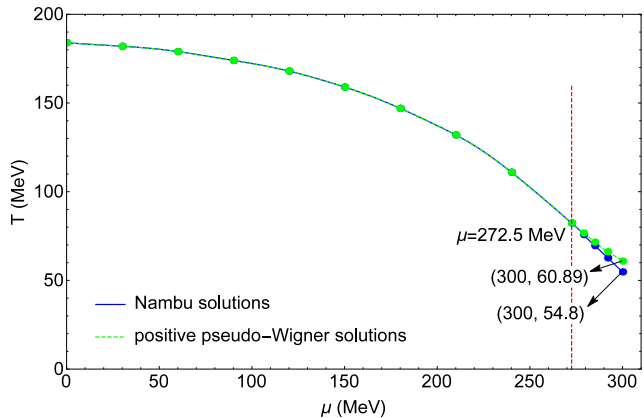}
\caption{Part of the QCD phase diagram obtained with our algorithm in this work. The blue solid line characterizes the vanishing edge of the Nambu solution, that is, the Nambu solution obtained with our algorithm exists under this line. The green dashed line denotes the starting edge of the positive pseudo-Wigner solution, i.e., the positive pseudo-Wigner solution obtained with our algorithm exists above this line. For $\mu\leq272.5$ MeV, these two lines coincide, implying the location of the crossover line in the QCD phase diagram. But for $\mu>272.5$ MeV, they separate from each other, causing a lack of information in the domain between these two lines.}
\label{Fig:phasediagram}
\end{figure}
Given the evolution process of the Nambu and pseudo-Wigner solution of the gap equation from the chiral limit to non-chiral limit in different schemes, we can get part of the QCD phase diagram with our algorithm, which is shown in Fig.~\ref{Fig:phasediagram}. In this figure, we denote the endpoint of the Nambu solution in the blue solid line, and the starting point of the positive pseudo-Wigner solution in the green dashed line. For $\mu<272.5$ MeV, these two lines coincide with each other, demonstrating a crossover between the corresponding two solutions, and the location of them just indicate a part of the crossover line predicted by our algorithm. But for $\mu>272.5$ MeV, these two lines separate from each other, presenting a lack of information of the QCD system in the separation area, thus our algorithm is invalid in this situation.

\section{SUMMARY AND DISCUSSION}\label{three}
In this paper, we briefly discuss some current perspectives on the chiral symmetry breaking and restoring at first, and then in the framework of the (2+1)-flavor NJL model, the solutions of the quark gap equation (especially the pseudo-Wigner solution) in the non-chiral limit are studied within a recently proposed algorithm. In this approach, supposing we have already obtained the solution $(M_{\rm u}(m_0),\, M_{\rm s}(m_0))$ of the quark gap equation in the case of $m=m_0$, then the derivatives $(M_{\rm u}',\, M_{\rm s}')$ only have one finite solution because of the linearity of the coupled derivative equations. Via the recursive formula of the effective quark masses, the value of $(M_{\rm u}(m_0+\Delta m),\, M_{\rm s}(m_0+\Delta m))$ ($\Delta m$ is a tiny variable) can be derived. Combining with the Nambu and Wigner solution $(M_{\rm u}(0),\, M_{\rm s}(0))$ in the chiral limit (convenient to calculate with a simple iteration method), we can get the Nambu and pseudo-Wigner solution $(M_{\rm u}(m),\, M_{\rm s}(m))$ for $m=5.5$ MeV in this work.

In particular, we apply this algorithm to two schemes: 1, $\mu=0$ and $T\neq0$; 2, $\mu\neq0$ and $T\neq0$. We find that the effective quark masses of the pseudo-Wigner solution in these two schemes both experience an absent region, while the Nambu solutions does not. By calculation and analysis, we find that for $\mu<272.5$ MeV where the second-order chiral phase transition happens as temperature increases in the chiral limit, things come similar to that for $\mu=0$ and $T\neq0$: the Nambu and positive pseudo-Wigner solution connect smoothly at a certain temperature, indicating a crossover in the QCD phase diagram. Furthermore, the solutions obtained with our algorithm are consistent with the strict physical solution obtained via the iterative method for $m=5.5$ MeV. However, for $\mu>272.5$ MeV, things are different: the Nambu solution does not connect to the positive pseudo-Wigner solution for any $m$, resulting in a missing of information of the QCD system in the discontinuous region and the invalidation of our algorithm.

For a better understanding of the evolution process of the Nambu and pseudo-Wigner solution with our algorithm, we present the chiral susceptibilities of the $u, d$ quark in these two schemes. We find that the diverging points of the chiral susceptibilities are same with the corresponding loss points of pseudo-Wigner solutions in the $m-T$ plane, because the derivatives $M_{\rm u}'$ and $\phi_{\rm u}'$ are tightly connected by Eq.~(\ref{condensatederiv}). In addition, the positive pseudo-Wigner solution does not encounter the loss of solutions for $\mu<272.5$ MeV, which is different from the scheme of $\mu>272.5$ MeV. For example, when $\mu=300$ MeV, the loss of solutions occur in the positive pseudo-Wigner solution for $m<1.62$ MeV, but in the negative pseudo-Wigner solution for $m>3.26$ MeV .

In conclusion, although the NJL model is a little rough as a phenomenological model of the QCD in some sense, it gives many qualitatively good predictions and explanations to QCD related fields such as the hadron physics. With the algorithm we employed in this paper, the evolution processes of the Nambu and pseudo-Wigner solution from the chiral limit to non-chiral limit can be clearly displayed. And for $\mu<272.5$ MeV in the QCD phase diagram, the Nambu and positive pseudo-Wigner solution we obtained are in good agreement with the strict physical solution of the gap equation, naturally giving a criterion of the starting point of the positive pseudo-Wigner solution.

In the future, a deeper understanding of the gluon propagator rather than taking it as a constant will improve the reliability of the model. In addition, the method beyond the mean field approximation might supply a more reliable calculation. Without doubt, it could not be better if the LQCD overcomes the "sign problem" and provides the most powerful simulations of the strong interacted matters at finite chemical potential.

\acknowledgments
This work is supported in part by the National Natural Science Foundation of China (under Grants No. 11475085, No. 11535005, No. 11690030, No. 11873030, and No. 11747140), the Fundamental Research Funds for the Central Universities (under Grant No. 020414380074), and by the National Major state Basic Research and Development of China (Grant No. 2016YFE0129300).

\bibliography{reference}

%merlin.mbs apsrev4-1.bst 2010-07-25 4.21a (PWD, AO, DPC) hacked
%Control: key (0)
%Control: author (8) initials jnrlst
%Control: editor formatted (1) identically to author
%Control: production of article title (-1) disabled
%Control: page (0) single
%Control: year (1) truncated
%Control: production of eprint (0) enabled
\begin{thebibliography}{53}%
\makeatletter
\providecommand \@ifxundefined [1]{%
 \@ifx{#1\undefined}
}%
\providecommand \@ifnum [1]{%
 \ifnum #1\expandafter \@firstoftwo
 \else \expandafter \@secondoftwo
 \fi
}%
\providecommand \@ifx [1]{%
 \ifx #1\expandafter \@firstoftwo
 \else \expandafter \@secondoftwo
 \fi
}%
\providecommand \natexlab [1]{#1}%
\providecommand \enquote  [1]{``#1''}%
\providecommand \bibnamefont  [1]{#1}%
\providecommand \bibfnamefont [1]{#1}%
\providecommand \citenamefont [1]{#1}%
\providecommand \href@noop [0]{\@secondoftwo}%
\providecommand \href [0]{\begingroup \@sanitize@url \@href}%
\providecommand \@href[1]{\@@startlink{#1}\@@href}%
\providecommand \@@href[1]{\endgroup#1\@@endlink}%
\providecommand \@sanitize@url [0]{\catcode `\\12\catcode `\$12\catcode
  `\&12\catcode `\#12\catcode `\^12\catcode `\_12\catcode `\%12\relax}%
\providecommand \@@startlink[1]{}%
\providecommand \@@endlink[0]{}%
\providecommand \url  [0]{\begingroup\@sanitize@url \@url }%
\providecommand \@url [1]{\endgroup\@href {#1}{\urlprefix }}%
\providecommand \urlprefix  [0]{URL }%
\providecommand \Eprint [0]{\href }%
\providecommand \doibase [0]{http://dx.doi.org/}%
\providecommand \selectlanguage [0]{\@gobble}%
\providecommand \bibinfo  [0]{\@secondoftwo}%
\providecommand \bibfield  [0]{\@secondoftwo}%
\providecommand \translation [1]{[#1]}%
\providecommand \BibitemOpen [0]{}%
\providecommand \bibitemStop [0]{}%
\providecommand \bibitemNoStop [0]{.\EOS\space}%
\providecommand \EOS [0]{\spacefactor3000\relax}%
\providecommand \BibitemShut  [1]{\csname bibitem#1\endcsname}%
\let\auto@bib@innerbib\@empty
%</preamble>
\bibitem [{\citenamefont {Bors{\'a}nyi}\ \emph {et~al.}(2010)\citenamefont
  {Bors{\'a}nyi}, \citenamefont {Fodor}, \citenamefont {Hoelbling},
  \citenamefont {Katz}, \citenamefont {Krieg}, \citenamefont {Ratti},\ and\
  \citenamefont {Szab{\'o}}}]{Borsanyi2010}%
  \BibitemOpen
  \bibfield  {author} {\bibinfo {author} {\bibfnamefont {S.}~\bibnamefont
  {Bors{\'a}nyi}}, \bibinfo {author} {\bibfnamefont {Z.}~\bibnamefont {Fodor}},
  \bibinfo {author} {\bibfnamefont {C.}~\bibnamefont {Hoelbling}}, \bibinfo
  {author} {\bibfnamefont {S.~D.}\ \bibnamefont {Katz}}, \bibinfo {author}
  {\bibfnamefont {S.}~\bibnamefont {Krieg}}, \bibinfo {author} {\bibfnamefont
  {C.}~\bibnamefont {Ratti}}, \ and\ \bibinfo {author} {\bibfnamefont {K.~K.}\
  \bibnamefont {Szab{\'o}}},\ }\href {\doibase 10.1007/JHEP09(2010)073}
  {\bibfield  {journal} {\bibinfo  {journal} {J. High Energy Phys.}\ }\textbf
  {\bibinfo {volume} {2010}},\ \bibinfo {pages} {73} (\bibinfo {year}
  {2010})}\BibitemShut {NoStop}%
\bibitem [{\citenamefont {Ejiri}\ and\ \citenamefont
  {Yamada}(2013)}]{PhysRevLett.110.172001}%
  \BibitemOpen
  \bibfield  {author} {\bibinfo {author} {\bibfnamefont {S.}~\bibnamefont
  {Ejiri}}\ and\ \bibinfo {author} {\bibfnamefont {N.}~\bibnamefont {Yamada}},\
  }\href {\doibase 10.1103/PhysRevLett.110.172001} {\bibfield  {journal}
  {\bibinfo  {journal} {Phys. Rev. Lett.}\ }\textbf {\bibinfo {volume} {110}},\
  \bibinfo {pages} {172001} (\bibinfo {year} {2013})}\BibitemShut {NoStop}%
\bibitem [{\citenamefont {Roberts}\ and\ \citenamefont
  {Williams}(1994)}]{ROBERTS1994477}%
  \BibitemOpen
  \bibfield  {author} {\bibinfo {author} {\bibfnamefont {C.~D.}\ \bibnamefont
  {Roberts}}\ and\ \bibinfo {author} {\bibfnamefont {A.~G.}\ \bibnamefont
  {Williams}},\ }\href {\doibase
  http://dx.doi.org/10.1016/0146-6410(94)90049-3} {\bibfield  {journal}
  {\bibinfo  {journal} {Prog. Part. Nucl. Phys.}\ }\textbf {\bibinfo {volume}
  {33}},\ \bibinfo {pages} {477 } (\bibinfo {year} {1994})}\BibitemShut
  {NoStop}%
\bibitem [{\citenamefont {Roberts}\ and\ \citenamefont
  {Schmidt}(2000)}]{Roberts2000S1}%
  \BibitemOpen
  \bibfield  {author} {\bibinfo {author} {\bibfnamefont {C.}~\bibnamefont
  {Roberts}}\ and\ \bibinfo {author} {\bibfnamefont {S.}~\bibnamefont
  {Schmidt}},\ }\href {\doibase
  http://dx.doi.org/10.1016/S0146-6410(00)90011-5} {\bibfield  {journal}
  {\bibinfo  {journal} {Prog. Part. Nucl. Phys.}\ }\textbf {\bibinfo {volume}
  {45, Supplement 1}},\ \bibinfo {pages} {S1 } (\bibinfo {year}
  {2000})}\BibitemShut {NoStop}%
\bibitem [{\citenamefont {Maris}\ and\ \citenamefont
  {Roberts}(2003)}]{doi:10.1142/S0218301303001326}%
  \BibitemOpen
  \bibfield  {author} {\bibinfo {author} {\bibfnamefont {P.}~\bibnamefont
  {Maris}}\ and\ \bibinfo {author} {\bibfnamefont {C.~D.}\ \bibnamefont
  {Roberts}},\ }\href {\doibase 10.1142/S0218301303001326} {\bibfield
  {journal} {\bibinfo  {journal} {Int. J. Mod. Phys. E}\ }\textbf {\bibinfo
  {volume} {12}},\ \bibinfo {pages} {297} (\bibinfo {year} {2003})}\BibitemShut
  {NoStop}%
\bibitem [{\citenamefont {Cl{\"o}et}\ and\ \citenamefont
  {Roberts}(2014)}]{Cloet20141}%
  \BibitemOpen
  \bibfield  {author} {\bibinfo {author} {\bibfnamefont {I.~C.}\ \bibnamefont
  {Cl{\"o}et}}\ and\ \bibinfo {author} {\bibfnamefont {C.~D.}\ \bibnamefont
  {Roberts}},\ }\href {\doibase http://dx.doi.org/10.1016/j.ppnp.2014.02.001}
  {\bibfield  {journal} {\bibinfo  {journal} {Prog. Part. Nucl. Phys.}\
  }\textbf {\bibinfo {volume} {77}},\ \bibinfo {pages} {1 } (\bibinfo {year}
  {2014})}\BibitemShut {NoStop}%
\bibitem [{\citenamefont {Zhao}\ \emph {et~al.}(2014)\citenamefont {Zhao},
  \citenamefont {Cui}, \citenamefont {Jiang},\ and\ \citenamefont
  {Zong}}]{PhysRevD.90.114031}%
  \BibitemOpen
  \bibfield  {author} {\bibinfo {author} {\bibfnamefont {A.-M.}\ \bibnamefont
  {Zhao}}, \bibinfo {author} {\bibfnamefont {Z.-F.}\ \bibnamefont {Cui}},
  \bibinfo {author} {\bibfnamefont {Y.}~\bibnamefont {Jiang}}, \ and\ \bibinfo
  {author} {\bibfnamefont {H.-S.}\ \bibnamefont {Zong}},\ }\href {\doibase
  10.1103/PhysRevD.90.114031} {\bibfield  {journal} {\bibinfo  {journal} {Phys.
  Rev. D}\ }\textbf {\bibinfo {volume} {90}},\ \bibinfo {pages} {114031}
  (\bibinfo {year} {2014})}\BibitemShut {NoStop}%
\bibitem [{\citenamefont {Xu}\ \emph {et~al.}(2015)\citenamefont {Xu},
  \citenamefont {Cui}, \citenamefont {Wang}, \citenamefont {Shi}, \citenamefont
  {Yang},\ and\ \citenamefont {Zong}}]{PhysRevD.91.056003}%
  \BibitemOpen
  \bibfield  {author} {\bibinfo {author} {\bibfnamefont {S.-S.}\ \bibnamefont
  {Xu}}, \bibinfo {author} {\bibfnamefont {Z.-F.}\ \bibnamefont {Cui}},
  \bibinfo {author} {\bibfnamefont {B.}~\bibnamefont {Wang}}, \bibinfo {author}
  {\bibfnamefont {Y.-M.}\ \bibnamefont {Shi}}, \bibinfo {author} {\bibfnamefont
  {Y.-C.}\ \bibnamefont {Yang}}, \ and\ \bibinfo {author} {\bibfnamefont
  {H.-S.}\ \bibnamefont {Zong}},\ }\href {\doibase 10.1103/PhysRevD.91.056003}
  {\bibfield  {journal} {\bibinfo  {journal} {Phys. Rev. D}\ }\textbf {\bibinfo
  {volume} {91}},\ \bibinfo {pages} {056003} (\bibinfo {year}
  {2015})}\BibitemShut {NoStop}%
\bibitem [{\citenamefont {Wang}\ \emph {et~al.}(2015)\citenamefont {Wang},
  \citenamefont {Wang}, \citenamefont {Cui},\ and\ \citenamefont
  {Zong}}]{PhysRevD.91.034017}%
  \BibitemOpen
  \bibfield  {author} {\bibinfo {author} {\bibfnamefont {B.}~\bibnamefont
  {Wang}}, \bibinfo {author} {\bibfnamefont {Y.-L.}\ \bibnamefont {Wang}},
  \bibinfo {author} {\bibfnamefont {Z.-F.}\ \bibnamefont {Cui}}, \ and\
  \bibinfo {author} {\bibfnamefont {H.-S.}\ \bibnamefont {Zong}},\ }\href
  {\doibase 10.1103/PhysRevD.91.034017} {\bibfield  {journal} {\bibinfo
  {journal} {Phys. Rev. D}\ }\textbf {\bibinfo {volume} {91}},\ \bibinfo
  {pages} {034017} (\bibinfo {year} {2015})}\BibitemShut {NoStop}%
\bibitem [{\citenamefont {Klevansky}(1992)}]{RevModPhys.64.649}%
  \BibitemOpen
  \bibfield  {author} {\bibinfo {author} {\bibfnamefont {S.~P.}\ \bibnamefont
  {Klevansky}},\ }\href {\doibase 10.1103/RevModPhys.64.649} {\bibfield
  {journal} {\bibinfo  {journal} {Rev. Mod. Phys.}\ }\textbf {\bibinfo {volume}
  {64}},\ \bibinfo {pages} {649} (\bibinfo {year} {1992})}\BibitemShut
  {NoStop}%
\bibitem [{\citenamefont {Buballa}(2005)}]{Buballa2005205}%
  \BibitemOpen
  \bibfield  {author} {\bibinfo {author} {\bibfnamefont {M.}~\bibnamefont
  {Buballa}},\ }\href {\doibase
  http://dx.doi.org/10.1016/j.physrep.2004.11.004} {\bibfield  {journal}
  {\bibinfo  {journal} {Phys. Rep.}\ }\textbf {\bibinfo {volume} {407}},\
  \bibinfo {pages} {205 } (\bibinfo {year} {2005})}\BibitemShut {NoStop}%
\bibitem [{\citenamefont {Cui}\ \emph {et~al.}(2013)\citenamefont {Cui},
  \citenamefont {Shi}, \citenamefont {Xia}, \citenamefont {Jiang},\ and\
  \citenamefont {Zong}}]{Cui2013}%
  \BibitemOpen
  \bibfield  {author} {\bibinfo {author} {\bibfnamefont {Z.-F.}\ \bibnamefont
  {Cui}}, \bibinfo {author} {\bibfnamefont {C.}~\bibnamefont {Shi}}, \bibinfo
  {author} {\bibfnamefont {Y.-H.}\ \bibnamefont {Xia}}, \bibinfo {author}
  {\bibfnamefont {Y.}~\bibnamefont {Jiang}}, \ and\ \bibinfo {author}
  {\bibfnamefont {H.-S.}\ \bibnamefont {Zong}},\ }\href {\doibase
  10.1140/epjc/s10052-013-2612-6} {\bibfield  {journal} {\bibinfo  {journal}
  {Eur. Phys. J. C}\ }\textbf {\bibinfo {volume} {73}},\ \bibinfo {pages}
  {2612} (\bibinfo {year} {2013})}\BibitemShut {NoStop}%
\bibitem [{\citenamefont {Kohyama}\ \emph
  {et~al.}(2015{\natexlab{a}})\citenamefont {Kohyama}, \citenamefont {Kimura},\
  and\ \citenamefont {Inagaki}}]{NuclPhysB.896.682}%
  \BibitemOpen
  \bibfield  {author} {\bibinfo {author} {\bibfnamefont {H.}~\bibnamefont
  {Kohyama}}, \bibinfo {author} {\bibfnamefont {D.}~\bibnamefont {Kimura}}, \
  and\ \bibinfo {author} {\bibfnamefont {T.}~\bibnamefont {Inagaki}},\
  }\href@noop {} {\bibfield  {journal} {\bibinfo  {journal} {Nucl. Phys. B}\
  }\textbf {\bibinfo {volume} {896}},\ \bibinfo {pages} {682} (\bibinfo {year}
  {2015}{\natexlab{a}})}\BibitemShut {NoStop}%
\bibitem [{\citenamefont {Fukushima}(2008{\natexlab{a}})}]{PhysRevD.78.039902}%
  \BibitemOpen
  \bibfield  {author} {\bibinfo {author} {\bibfnamefont {K.}~\bibnamefont
  {Fukushima}},\ }\href {\doibase 10.1103/PhysRevD.78.039902} {\bibfield
  {journal} {\bibinfo  {journal} {Phys. Rev. D}\ }\textbf {\bibinfo {volume}
  {78}},\ \bibinfo {pages} {039902} (\bibinfo {year}
  {2008}{\natexlab{a}})}\BibitemShut {NoStop}%
\bibitem [{\citenamefont {Boeckel}\ and\ \citenamefont
  {Schaffner-Bielich}(2010)}]{PhysRevLett.105.041301}%
  \BibitemOpen
  \bibfield  {author} {\bibinfo {author} {\bibfnamefont {T.}~\bibnamefont
  {Boeckel}}\ and\ \bibinfo {author} {\bibfnamefont {J.}~\bibnamefont
  {Schaffner-Bielich}},\ }\href {\doibase 10.1103/PhysRevLett.105.041301}
  {\bibfield  {journal} {\bibinfo  {journal} {Phys. Rev. Lett.}\ }\textbf
  {\bibinfo {volume} {105}},\ \bibinfo {pages} {041301} (\bibinfo {year}
  {2010})}\BibitemShut {NoStop}%
\bibitem [{\citenamefont {Enqvist}\ and\ \citenamefont
  {Sirkka}(1993)}]{ENQVIST1993298}%
  \BibitemOpen
  \bibfield  {author} {\bibinfo {author} {\bibfnamefont {K.}~\bibnamefont
  {Enqvist}}\ and\ \bibinfo {author} {\bibfnamefont {J.}~\bibnamefont
  {Sirkka}},\ }\href {\doibase https://doi.org/10.1016/0370-2693(93)91239-J}
  {\bibfield  {journal} {\bibinfo  {journal} {Physics Letters B}\ }\textbf
  {\bibinfo {volume} {314}},\ \bibinfo {pages} {298 } (\bibinfo {year}
  {1993})}\BibitemShut {NoStop}%
\bibitem [{\citenamefont {{Ai}}\ \emph {et~al.}(2018)\citenamefont {{Ai}},
  \citenamefont {{Gao}}, \citenamefont {{Dai}}, \citenamefont {{Wu}},
  \citenamefont {{Li}}, \citenamefont {{Zhang}},\ and\ \citenamefont
  {{Li}}}]{2018ApJ...860...57A}%
  \BibitemOpen
  \bibfield  {author} {\bibinfo {author} {\bibfnamefont {S.}~\bibnamefont
  {{Ai}}}, \bibinfo {author} {\bibfnamefont {H.}~\bibnamefont {{Gao}}},
  \bibinfo {author} {\bibfnamefont {Z.-G.}\ \bibnamefont {{Dai}}}, \bibinfo
  {author} {\bibfnamefont {X.-F.}\ \bibnamefont {{Wu}}}, \bibinfo {author}
  {\bibfnamefont {A.}~\bibnamefont {{Li}}}, \bibinfo {author} {\bibfnamefont
  {B.}~\bibnamefont {{Zhang}}}, \ and\ \bibinfo {author} {\bibfnamefont
  {M.-Z.}\ \bibnamefont {{Li}}},\ }\href {\doibase 10.3847/1538-4357/aac2b7}
  {\bibfield  {journal} {\bibinfo  {journal} {Astrophys. J}\ }\textbf {\bibinfo
  {volume} {860}},\ \bibinfo {eid} {57} (\bibinfo {year} {2018})},\ \Eprint
  {http://arxiv.org/abs/1802.00571} {arXiv:1802.00571} \BibitemShut {NoStop}%
\bibitem [{\citenamefont {Li}\ \emph {et~al.}(2018{\natexlab{a}})\citenamefont
  {Li}, \citenamefont {Yan}, \citenamefont {Geng}, \citenamefont {Huang},\ and\
  \citenamefont {Zong}}]{PhysRevD.98.083013}%
  \BibitemOpen
  \bibfield  {author} {\bibinfo {author} {\bibfnamefont {C.-M.}\ \bibnamefont
  {Li}}, \bibinfo {author} {\bibfnamefont {Y.}~\bibnamefont {Yan}}, \bibinfo
  {author} {\bibfnamefont {J.-J.}\ \bibnamefont {Geng}}, \bibinfo {author}
  {\bibfnamefont {Y.-F.}\ \bibnamefont {Huang}}, \ and\ \bibinfo {author}
  {\bibfnamefont {H.-S.}\ \bibnamefont {Zong}},\ }\href {\doibase
  10.1103/PhysRevD.98.083013} {\bibfield  {journal} {\bibinfo  {journal} {Phys.
  Rev. D}\ }\textbf {\bibinfo {volume} {98}},\ \bibinfo {pages} {083013}
  (\bibinfo {year} {2018}{\natexlab{a}})}\BibitemShut {NoStop}%
\bibitem [{\citenamefont {Li}\ \emph {et~al.}(2018{\natexlab{b}})\citenamefont
  {Li}, \citenamefont {Zhang}, \citenamefont {Yan}, \citenamefont {Huang},\
  and\ \citenamefont {Zong}}]{PhysRevD.97.103013}%
  \BibitemOpen
  \bibfield  {author} {\bibinfo {author} {\bibfnamefont {C.-M.}\ \bibnamefont
  {Li}}, \bibinfo {author} {\bibfnamefont {J.-L.}\ \bibnamefont {Zhang}},
  \bibinfo {author} {\bibfnamefont {Y.}~\bibnamefont {Yan}}, \bibinfo {author}
  {\bibfnamefont {Y.-F.}\ \bibnamefont {Huang}}, \ and\ \bibinfo {author}
  {\bibfnamefont {H.-S.}\ \bibnamefont {Zong}},\ }\href {\doibase
  10.1103/PhysRevD.97.103013} {\bibfield  {journal} {\bibinfo  {journal} {Phys.
  Rev. D}\ }\textbf {\bibinfo {volume} {97}},\ \bibinfo {pages} {103013}
  (\bibinfo {year} {2018}{\natexlab{b}})}\BibitemShut {NoStop}%
\bibitem [{\citenamefont {Li}\ \emph {et~al.}(2017)\citenamefont {Li},
  \citenamefont {Zhang}, \citenamefont {Zhao}, \citenamefont {Zhao},\ and\
  \citenamefont {Zong}}]{PhysRevD.95.056018}%
  \BibitemOpen
  \bibfield  {author} {\bibinfo {author} {\bibfnamefont {C.-M.}\ \bibnamefont
  {Li}}, \bibinfo {author} {\bibfnamefont {J.-L.}\ \bibnamefont {Zhang}},
  \bibinfo {author} {\bibfnamefont {T.}~\bibnamefont {Zhao}}, \bibinfo {author}
  {\bibfnamefont {Y.-P.}\ \bibnamefont {Zhao}}, \ and\ \bibinfo {author}
  {\bibfnamefont {H.-S.}\ \bibnamefont {Zong}},\ }\href {\doibase
  10.1103/PhysRevD.95.056018} {\bibfield  {journal} {\bibinfo  {journal} {Phys.
  Rev. D}\ }\textbf {\bibinfo {volume} {95}},\ \bibinfo {pages} {056018}
  (\bibinfo {year} {2017})}\BibitemShut {NoStop}%
\bibitem [{\citenamefont {Yan}\ \emph {et~al.}(2012)\citenamefont {Yan},
  \citenamefont {Cao}, \citenamefont {Luo}, \citenamefont {Sun},\ and\
  \citenamefont {Zong}}]{PhysRevD.86.114028}%
  \BibitemOpen
  \bibfield  {author} {\bibinfo {author} {\bibfnamefont {Y.}~\bibnamefont
  {Yan}}, \bibinfo {author} {\bibfnamefont {J.}~\bibnamefont {Cao}}, \bibinfo
  {author} {\bibfnamefont {X.-L.}\ \bibnamefont {Luo}}, \bibinfo {author}
  {\bibfnamefont {W.-M.}\ \bibnamefont {Sun}}, \ and\ \bibinfo {author}
  {\bibfnamefont {H.-S.}\ \bibnamefont {Zong}},\ }\href {\doibase
  10.1103/PhysRevD.86.114028} {\bibfield  {journal} {\bibinfo  {journal} {Phys.
  Rev. D}\ }\textbf {\bibinfo {volume} {86}},\ \bibinfo {pages} {114028}
  (\bibinfo {year} {2012})}\BibitemShut {NoStop}%
\bibitem [{\citenamefont {Wu}\ and\ \citenamefont
  {Ren}(2011)}]{PhysRevC.83.025805}%
  \BibitemOpen
  \bibfield  {author} {\bibinfo {author} {\bibfnamefont {C.}~\bibnamefont
  {Wu}}\ and\ \bibinfo {author} {\bibfnamefont {Z.}~\bibnamefont {Ren}},\
  }\href {\doibase 10.1103/PhysRevC.83.025805} {\bibfield  {journal} {\bibinfo
  {journal} {Phys. Rev. C}\ }\textbf {\bibinfo {volume} {83}},\ \bibinfo
  {pages} {025805} (\bibinfo {year} {2011})}\BibitemShut {NoStop}%
\bibitem [{\citenamefont {WU}\ \emph {et~al.}(2013)\citenamefont {WU},
  \citenamefont {QIAN}, \citenamefont {MA},\ and\ \citenamefont
  {YANG}}]{doi:10.1142/S0218301313500262}%
  \BibitemOpen
  \bibfield  {author} {\bibinfo {author} {\bibfnamefont {C.}~\bibnamefont
  {WU}}, \bibinfo {author} {\bibfnamefont {W.-L.}\ \bibnamefont {QIAN}},
  \bibinfo {author} {\bibfnamefont {Y.-G.}\ \bibnamefont {MA}}, \ and\ \bibinfo
  {author} {\bibfnamefont {J.-F.}\ \bibnamefont {YANG}},\ }\href {\doibase
  10.1142/S0218301313500262} {\bibfield  {journal} {\bibinfo  {journal} {Int.
  J. Mod. Phys. E}\ }\textbf {\bibinfo {volume} {22}},\ \bibinfo {pages}
  {1350026} (\bibinfo {year} {2013})},\ \Eprint
  {http://arxiv.org/abs/https://doi.org/10.1142/S0218301313500262}
  {https://doi.org/10.1142/S0218301313500262} \BibitemShut {NoStop}%
\bibitem [{\citenamefont {Pan}\ \emph {et~al.}(2017)\citenamefont {Pan},
  \citenamefont {Cui}, \citenamefont {Chang},\ and\ \citenamefont
  {Zong}}]{doi:10.1142/S0217751X17500671}%
  \BibitemOpen
  \bibfield  {author} {\bibinfo {author} {\bibfnamefont {Z.}~\bibnamefont
  {Pan}}, \bibinfo {author} {\bibfnamefont {Z.-F.}\ \bibnamefont {Cui}},
  \bibinfo {author} {\bibfnamefont {C.-H.}\ \bibnamefont {Chang}}, \ and\
  \bibinfo {author} {\bibfnamefont {H.-S.}\ \bibnamefont {Zong}},\ }\href
  {\doibase 10.1142/S0217751X17500671} {\bibfield  {journal} {\bibinfo
  {journal} {Int. J. Mod. Phys. A}\ }\textbf {\bibinfo {volume} {32}},\
  \bibinfo {pages} {1750067} (\bibinfo {year} {2017})},\ \Eprint
  {http://arxiv.org/abs/https://doi.org/10.1142/S0217751X17500671}
  {https://doi.org/10.1142/S0217751X17500671} \BibitemShut {NoStop}%
\bibitem [{\citenamefont {Cui}\ \emph {et~al.}(2017)\citenamefont {Cui},
  \citenamefont {Zhang},\ and\ \citenamefont {Zong}}]{10.1038/srep45937}%
  \BibitemOpen
  \bibfield  {author} {\bibinfo {author} {\bibfnamefont {Z.-F.}\ \bibnamefont
  {Cui}}, \bibinfo {author} {\bibfnamefont {J.-L.}\ \bibnamefont {Zhang}}, \
  and\ \bibinfo {author} {\bibfnamefont {H.-S.}\ \bibnamefont {Zong}},\ }\href
  {\doibase 10.1038/srep45937} {\bibfield  {journal} {\bibinfo  {journal} {Sci.
  Rep.}\ }\textbf {\bibinfo {volume} {7}},\ \bibinfo {pages} {45937} (\bibinfo
  {year} {2017})}\BibitemShut {NoStop}%
\bibitem [{\citenamefont {Kov\'acs}\ \emph {et~al.}(2016)\citenamefont
  {Kov\'acs}, \citenamefont {Sz\'ep},\ and\ \citenamefont
  {Wolf}}]{PhysRevD.93.114014}%
  \BibitemOpen
  \bibfield  {author} {\bibinfo {author} {\bibfnamefont {P.}~\bibnamefont
  {Kov\'acs}}, \bibinfo {author} {\bibfnamefont {Z.}~\bibnamefont {Sz\'ep}}, \
  and\ \bibinfo {author} {\bibfnamefont {G.}~\bibnamefont {Wolf}},\ }\href
  {\doibase 10.1103/PhysRevD.93.114014} {\bibfield  {journal} {\bibinfo
  {journal} {Phys. Rev. D}\ }\textbf {\bibinfo {volume} {93}},\ \bibinfo
  {pages} {114014} (\bibinfo {year} {2016})}\BibitemShut {NoStop}%
\bibitem [{\citenamefont {Eichmann}\ \emph {et~al.}(2016)\citenamefont
  {Eichmann}, \citenamefont {Fischer},\ and\ \citenamefont
  {Welzbacher}}]{PhysRevD.93.034013}%
  \BibitemOpen
  \bibfield  {author} {\bibinfo {author} {\bibfnamefont {G.}~\bibnamefont
  {Eichmann}}, \bibinfo {author} {\bibfnamefont {C.~S.}\ \bibnamefont
  {Fischer}}, \ and\ \bibinfo {author} {\bibfnamefont {C.~A.}\ \bibnamefont
  {Welzbacher}},\ }\href {\doibase 10.1103/PhysRevD.93.034013} {\bibfield
  {journal} {\bibinfo  {journal} {Phys. Rev. D}\ }\textbf {\bibinfo {volume}
  {93}},\ \bibinfo {pages} {034013} (\bibinfo {year} {2016})}\BibitemShut
  {NoStop}%
\bibitem [{\citenamefont {Ayala}\ \emph {et~al.}(2015)\citenamefont {Ayala},
  \citenamefont {Bashir}, \citenamefont {Cobos-Mart¨ªnez}, \citenamefont
  {Hern¨¢ndez-Ortiz},\ and\ \citenamefont {Raya}}]{AYALA201577}%
  \BibitemOpen
  \bibfield  {author} {\bibinfo {author} {\bibfnamefont {A.}~\bibnamefont
  {Ayala}}, \bibinfo {author} {\bibfnamefont {A.}~\bibnamefont {Bashir}},
  \bibinfo {author} {\bibfnamefont {J.}~\bibnamefont {Cobos-Mart¨ªnez}},
  \bibinfo {author} {\bibfnamefont {S.}~\bibnamefont {Hern¨¢ndez-Ortiz}}, \
  and\ \bibinfo {author} {\bibfnamefont {A.}~\bibnamefont {Raya}},\ }\href
  {\doibase https://doi.org/10.1016/j.nuclphysb.2015.05.014} {\bibfield
  {journal} {\bibinfo  {journal} {Nucl. Phys. B}\ }\textbf {\bibinfo {volume}
  {897}},\ \bibinfo {pages} {77 } (\bibinfo {year} {2015})}\BibitemShut
  {NoStop}%
\bibitem [{\citenamefont {Kohyama}\ \emph
  {et~al.}(2015{\natexlab{b}})\citenamefont {Kohyama}, \citenamefont {Kimura},\
  and\ \citenamefont {Inagaki}}]{KOHYAMA2015682}%
  \BibitemOpen
  \bibfield  {author} {\bibinfo {author} {\bibfnamefont {H.}~\bibnamefont
  {Kohyama}}, \bibinfo {author} {\bibfnamefont {D.}~\bibnamefont {Kimura}}, \
  and\ \bibinfo {author} {\bibfnamefont {T.}~\bibnamefont {Inagaki}},\ }\href
  {\doibase https://doi.org/10.1016/j.nuclphysb.2015.05.015} {\bibfield
  {journal} {\bibinfo  {journal} {Nucl. Phys. B}\ }\textbf {\bibinfo {volume}
  {896}},\ \bibinfo {pages} {682 } (\bibinfo {year}
  {2015}{\natexlab{b}})}\BibitemShut {NoStop}%
\bibitem [{\citenamefont {Marquez}\ \emph {et~al.}(2015)\citenamefont
  {Marquez}, \citenamefont {Ahmad}, \citenamefont {Buballa},\ and\
  \citenamefont {Raya}}]{MARQUEZ2015529}%
  \BibitemOpen
  \bibfield  {author} {\bibinfo {author} {\bibfnamefont {F.}~\bibnamefont
  {Marquez}}, \bibinfo {author} {\bibfnamefont {A.}~\bibnamefont {Ahmad}},
  \bibinfo {author} {\bibfnamefont {M.}~\bibnamefont {Buballa}}, \ and\
  \bibinfo {author} {\bibfnamefont {A.}~\bibnamefont {Raya}},\ }\href {\doibase
  https://doi.org/10.1016/j.physletb.2015.06.031} {\bibfield  {journal}
  {\bibinfo  {journal} {Phys. Lett. B}\ }\textbf {\bibinfo {volume} {747}},\
  \bibinfo {pages} {529 } (\bibinfo {year} {2015})}\BibitemShut {NoStop}%
\bibitem [{\citenamefont {Cui}\ \emph {et~al.}(2018)\citenamefont {Cui},
  \citenamefont {Xu}, \citenamefont {Li}, \citenamefont {Sun}, \citenamefont
  {Zhang},\ and\ \citenamefont {Zong}}]{Cui2018}%
  \BibitemOpen
  \bibfield  {author} {\bibinfo {author} {\bibfnamefont {Z.-F.}\ \bibnamefont
  {Cui}}, \bibinfo {author} {\bibfnamefont {S.-S.}\ \bibnamefont {Xu}},
  \bibinfo {author} {\bibfnamefont {B.-L.}\ \bibnamefont {Li}}, \bibinfo
  {author} {\bibfnamefont {A.}~\bibnamefont {Sun}}, \bibinfo {author}
  {\bibfnamefont {J.-B.}\ \bibnamefont {Zhang}}, \ and\ \bibinfo {author}
  {\bibfnamefont {H.-S.}\ \bibnamefont {Zong}},\ }\href {\doibase
  10.1140/epjc/s10052-018-6264-4} {\bibfield  {journal} {\bibinfo  {journal}
  {Eur. Phys. J. C}\ }\textbf {\bibinfo {volume} {78}},\ \bibinfo {pages} {770}
  (\bibinfo {year} {2018})}\BibitemShut {NoStop}%
\bibitem [{\citenamefont {Schaefer}\ \emph {et~al.}(2007)\citenamefont
  {Schaefer}, \citenamefont {Pawlowski},\ and\ \citenamefont
  {Wambach}}]{PhysRevD.76.074023}%
  \BibitemOpen
  \bibfield  {author} {\bibinfo {author} {\bibfnamefont {B.-J.}\ \bibnamefont
  {Schaefer}}, \bibinfo {author} {\bibfnamefont {J.~M.}\ \bibnamefont
  {Pawlowski}}, \ and\ \bibinfo {author} {\bibfnamefont {J.}~\bibnamefont
  {Wambach}},\ }\href {\doibase 10.1103/PhysRevD.76.074023} {\bibfield
  {journal} {\bibinfo  {journal} {Phys. Rev. D}\ }\textbf {\bibinfo {volume}
  {76}},\ \bibinfo {pages} {074023} (\bibinfo {year} {2007})}\BibitemShut
  {NoStop}%
\bibitem [{\citenamefont {Kl{\"a}hn}\ \emph {et~al.}(2017)\citenamefont
  {Kl{\"a}hn}, \citenamefont {Fischer},\ and\ \citenamefont
  {Hempel}}]{0004-637X-836-1-89}%
  \BibitemOpen
  \bibfield  {author} {\bibinfo {author} {\bibfnamefont {T.}~\bibnamefont
  {Kl{\"a}hn}}, \bibinfo {author} {\bibfnamefont {T.}~\bibnamefont {Fischer}},
  \ and\ \bibinfo {author} {\bibfnamefont {M.}~\bibnamefont {Hempel}},\ }\href
  {http://stacks.iop.org/0004-637X/836/i=1/a=89} {\bibfield  {journal}
  {\bibinfo  {journal} {Astrophys. J}\ }\textbf {\bibinfo {volume} {836}},\
  \bibinfo {pages} {89} (\bibinfo {year} {2017})}\BibitemShut {NoStop}%
\bibitem [{\citenamefont {Zong}\ \emph {et~al.}(2005)\citenamefont {Zong},
  \citenamefont {Sun}, \citenamefont {Ping}, \citenamefont {Lu},\ and\
  \citenamefont {Wang}}]{0256-307X-22-12-014}%
  \BibitemOpen
  \bibfield  {author} {\bibinfo {author} {\bibfnamefont {H.-S.}\ \bibnamefont
  {Zong}}, \bibinfo {author} {\bibfnamefont {W.-M.}\ \bibnamefont {Sun}},
  \bibinfo {author} {\bibfnamefont {J.-L.}\ \bibnamefont {Ping}}, \bibinfo
  {author} {\bibfnamefont {X.-F.}\ \bibnamefont {Lu}}, \ and\ \bibinfo {author}
  {\bibfnamefont {F.}~\bibnamefont {Wang}},\ }\href
  {http://stacks.iop.org/0256-307X/22/i=12/a=014} {\bibfield  {journal}
  {\bibinfo  {journal} {Chin. Phys. Lett.}\ }\textbf {\bibinfo {volume} {22}},\
  \bibinfo {pages} {3036} (\bibinfo {year} {2005})}\BibitemShut {NoStop}%
\bibitem [{\citenamefont {Chang}\ \emph {et~al.}(2007)\citenamefont {Chang},
  \citenamefont {Liu}, \citenamefont {Bhagwat}, \citenamefont {Roberts},\ and\
  \citenamefont {Wright}}]{PhysRevC.75.015201}%
  \BibitemOpen
  \bibfield  {author} {\bibinfo {author} {\bibfnamefont {L.}~\bibnamefont
  {Chang}}, \bibinfo {author} {\bibfnamefont {Y.-X.}\ \bibnamefont {Liu}},
  \bibinfo {author} {\bibfnamefont {M.~S.}\ \bibnamefont {Bhagwat}}, \bibinfo
  {author} {\bibfnamefont {C.~D.}\ \bibnamefont {Roberts}}, \ and\ \bibinfo
  {author} {\bibfnamefont {S.~V.}\ \bibnamefont {Wright}},\ }\href {\doibase
  10.1103/PhysRevC.75.015201} {\bibfield  {journal} {\bibinfo  {journal} {Phys.
  Rev. C}\ }\textbf {\bibinfo {volume} {75}},\ \bibinfo {pages} {015201}
  (\bibinfo {year} {2007})}\BibitemShut {NoStop}%
\bibitem [{\citenamefont {Wang}\ \emph {et~al.}(2012)\citenamefont {Wang},
  \citenamefont {Qin}, \citenamefont {Liu}, \citenamefont {Chang},
  \citenamefont {Roberts},\ and\ \citenamefont {Schmidt}}]{PhysRevD.86.114001}%
  \BibitemOpen
  \bibfield  {author} {\bibinfo {author} {\bibfnamefont {K.-L.}\ \bibnamefont
  {Wang}}, \bibinfo {author} {\bibfnamefont {S.-X.}\ \bibnamefont {Qin}},
  \bibinfo {author} {\bibfnamefont {Y.-X.}\ \bibnamefont {Liu}}, \bibinfo
  {author} {\bibfnamefont {L.}~\bibnamefont {Chang}}, \bibinfo {author}
  {\bibfnamefont {C.~D.}\ \bibnamefont {Roberts}}, \ and\ \bibinfo {author}
  {\bibfnamefont {S.~M.}\ \bibnamefont {Schmidt}},\ }\href {\doibase
  10.1103/PhysRevD.86.114001} {\bibfield  {journal} {\bibinfo  {journal} {Phys.
  Rev. D}\ }\textbf {\bibinfo {volume} {86}},\ \bibinfo {pages} {114001}
  (\bibinfo {year} {2012})}\BibitemShut {NoStop}%
\bibitem [{\citenamefont {Williams}\ \emph {et~al.}(2007)\citenamefont
  {Williams}, \citenamefont {Fischer},\ and\ \citenamefont
  {Pennington}}]{WILLIAMS2007167}%
  \BibitemOpen
  \bibfield  {author} {\bibinfo {author} {\bibfnamefont {R.}~\bibnamefont
  {Williams}}, \bibinfo {author} {\bibfnamefont {C.}~\bibnamefont {Fischer}}, \
  and\ \bibinfo {author} {\bibfnamefont {M.}~\bibnamefont {Pennington}},\
  }\href {\doibase https://doi.org/10.1016/j.physletb.2006.12.055} {\bibfield
  {journal} {\bibinfo  {journal} {Phys. Lett. B}\ }\textbf {\bibinfo {volume}
  {645}},\ \bibinfo {pages} {167 } (\bibinfo {year} {2007})}\BibitemShut
  {NoStop}%
\bibitem [{\citenamefont {Cui}\ \emph {et~al.}(2014)\citenamefont {Cui},
  \citenamefont {Shi}, \citenamefont {Sun}, \citenamefont {Wang},\ and\
  \citenamefont {Zong}}]{Cui2014}%
  \BibitemOpen
  \bibfield  {author} {\bibinfo {author} {\bibfnamefont {Z.-F.}\ \bibnamefont
  {Cui}}, \bibinfo {author} {\bibfnamefont {C.}~\bibnamefont {Shi}}, \bibinfo
  {author} {\bibfnamefont {W.-M.}\ \bibnamefont {Sun}}, \bibinfo {author}
  {\bibfnamefont {Y.-L.}\ \bibnamefont {Wang}}, \ and\ \bibinfo {author}
  {\bibfnamefont {H.-S.}\ \bibnamefont {Zong}},\ }\href {\doibase
  10.1140/epjc/s10052-014-2782-x} {\bibfield  {journal} {\bibinfo  {journal}
  {Eur. Phys. J. C}\ }\textbf {\bibinfo {volume} {74}},\ \bibinfo {pages}
  {2782} (\bibinfo {year} {2014})}\BibitemShut {NoStop}%
\bibitem [{\citenamefont {Raya}\ \emph {et~al.}(2013)\citenamefont {Raya},
  \citenamefont {Bashir}, \citenamefont {Hern\'andez-Ortiz}, \citenamefont
  {Raya},\ and\ \citenamefont {Roberts}}]{PhysRevD.88.096003}%
  \BibitemOpen
  \bibfield  {author} {\bibinfo {author} {\bibfnamefont {K.}~\bibnamefont
  {Raya}}, \bibinfo {author} {\bibfnamefont {A.}~\bibnamefont {Bashir}},
  \bibinfo {author} {\bibfnamefont {S.}~\bibnamefont {Hern\'andez-Ortiz}},
  \bibinfo {author} {\bibfnamefont {A.}~\bibnamefont {Raya}}, \ and\ \bibinfo
  {author} {\bibfnamefont {C.~D.}\ \bibnamefont {Roberts}},\ }\href {\doibase
  10.1103/PhysRevD.88.096003} {\bibfield  {journal} {\bibinfo  {journal} {Phys.
  Rev. D}\ }\textbf {\bibinfo {volume} {88}},\ \bibinfo {pages} {096003}
  (\bibinfo {year} {2013})}\BibitemShut {NoStop}%
\bibitem [{\citenamefont {Wang}\ \emph {et~al.}(2016)\citenamefont {Wang},
  \citenamefont {Cui},\ and\ \citenamefont {Zong}}]{PhysRevD.94.096003}%
  \BibitemOpen
  \bibfield  {author} {\bibinfo {author} {\bibfnamefont {Q.-W.}\ \bibnamefont
  {Wang}}, \bibinfo {author} {\bibfnamefont {Z.-F.}\ \bibnamefont {Cui}}, \
  and\ \bibinfo {author} {\bibfnamefont {H.-S.}\ \bibnamefont {Zong}},\ }\href
  {\doibase 10.1103/PhysRevD.94.096003} {\bibfield  {journal} {\bibinfo
  {journal} {Phys. Rev. D}\ }\textbf {\bibinfo {volume} {94}},\ \bibinfo
  {pages} {096003} (\bibinfo {year} {2016})}\BibitemShut {NoStop}%
\bibitem [{\citenamefont {Maris}\ \emph {et~al.}(2001)\citenamefont {Maris},
  \citenamefont {Roberts}, \citenamefont {Schmidt},\ and\ \citenamefont
  {Tandy}}]{PhysRevC.63.025202}%
  \BibitemOpen
  \bibfield  {author} {\bibinfo {author} {\bibfnamefont {P.}~\bibnamefont
  {Maris}}, \bibinfo {author} {\bibfnamefont {C.~D.}\ \bibnamefont {Roberts}},
  \bibinfo {author} {\bibfnamefont {S.~M.}\ \bibnamefont {Schmidt}}, \ and\
  \bibinfo {author} {\bibfnamefont {P.~C.}\ \bibnamefont {Tandy}},\ }\href
  {\doibase 10.1103/PhysRevC.63.025202} {\bibfield  {journal} {\bibinfo
  {journal} {Phys. Rev. C}\ }\textbf {\bibinfo {volume} {63}},\ \bibinfo
  {pages} {025202} (\bibinfo {year} {2001})}\BibitemShut {NoStop}%
\bibitem [{\citenamefont {Jiang}\ \emph {et~al.}(2012)\citenamefont {Jiang},
  \citenamefont {Gong}, \citenamefont {Sun},\ and\ \citenamefont
  {Zong}}]{0256-307X-29-4-041201}%
  \BibitemOpen
  \bibfield  {author} {\bibinfo {author} {\bibfnamefont {Y.}~\bibnamefont
  {Jiang}}, \bibinfo {author} {\bibfnamefont {H.}~\bibnamefont {Gong}},
  \bibinfo {author} {\bibfnamefont {W.-M.}\ \bibnamefont {Sun}}, \ and\
  \bibinfo {author} {\bibfnamefont {H.-S.}\ \bibnamefont {Zong}},\ }\href
  {http://stacks.iop.org/0256-307X/29/i=4/a=041201} {\bibfield  {journal}
  {\bibinfo  {journal} {Chin. Phys. Lett.}\ }\textbf {\bibinfo {volume} {29}},\
  \bibinfo {pages} {041201} (\bibinfo {year} {2012})}\BibitemShut {NoStop}%
\bibitem [{\citenamefont {Qin}\ \emph {et~al.}(2011)\citenamefont {Qin},
  \citenamefont {Chang}, \citenamefont {Chen}, \citenamefont {Liu},\ and\
  \citenamefont {Roberts}}]{PhysRevLett.106.172301}%
  \BibitemOpen
  \bibfield  {author} {\bibinfo {author} {\bibfnamefont {S.-X.}\ \bibnamefont
  {Qin}}, \bibinfo {author} {\bibfnamefont {L.}~\bibnamefont {Chang}}, \bibinfo
  {author} {\bibfnamefont {H.}~\bibnamefont {Chen}}, \bibinfo {author}
  {\bibfnamefont {Y.-X.}\ \bibnamefont {Liu}}, \ and\ \bibinfo {author}
  {\bibfnamefont {C.~D.}\ \bibnamefont {Roberts}},\ }\href {\doibase
  10.1103/PhysRevLett.106.172301} {\bibfield  {journal} {\bibinfo  {journal}
  {Phys. Rev. Lett.}\ }\textbf {\bibinfo {volume} {106}},\ \bibinfo {pages}
  {172301} (\bibinfo {year} {2011})}\BibitemShut {NoStop}%
\bibitem [{\citenamefont {Xu}\ \emph {et~al.}(2018)\citenamefont {Xu},
  \citenamefont {Cui}, \citenamefont {Sun},\ and\ \citenamefont
  {Zong}}]{0954-3899-45-10-105001}%
  \BibitemOpen
  \bibfield  {author} {\bibinfo {author} {\bibfnamefont {S.-S.}\ \bibnamefont
  {Xu}}, \bibinfo {author} {\bibfnamefont {Z.-F.}\ \bibnamefont {Cui}},
  \bibinfo {author} {\bibfnamefont {A.}~\bibnamefont {Sun}}, \ and\ \bibinfo
  {author} {\bibfnamefont {H.-S.}\ \bibnamefont {Zong}},\ }\href
  {http://stacks.iop.org/0954-3899/45/i=10/a=105001} {\bibfield  {journal}
  {\bibinfo  {journal} {J. Phys. G: Nucl. Part. Phys.}\ }\textbf {\bibinfo
  {volume} {45}},\ \bibinfo {pages} {105001} (\bibinfo {year}
  {2018})}\BibitemShut {NoStop}%
\bibitem [{\citenamefont {Hatsuda}\ and\ \citenamefont
  {Kunihiro}(1994)}]{HATSUDA1994221}%
  \BibitemOpen
  \bibfield  {author} {\bibinfo {author} {\bibfnamefont {T.}~\bibnamefont
  {Hatsuda}}\ and\ \bibinfo {author} {\bibfnamefont {T.}~\bibnamefont
  {Kunihiro}},\ }\href {\doibase
  http://dx.doi.org/10.1016/0370-1573(94)90022-1} {\bibfield  {journal}
  {\bibinfo  {journal} {Phys. Rep.}\ }\textbf {\bibinfo {volume} {247}},\
  \bibinfo {pages} {221 } (\bibinfo {year} {1994})}\BibitemShut {NoStop}%
\bibitem [{\citenamefont {Fukushima}(2008{\natexlab{b}})}]{PhysRevD.77.114028}%
  \BibitemOpen
  \bibfield  {author} {\bibinfo {author} {\bibfnamefont {K.}~\bibnamefont
  {Fukushima}},\ }\href {\doibase 10.1103/PhysRevD.77.114028} {\bibfield
  {journal} {\bibinfo  {journal} {Phys. Rev. D}\ }\textbf {\bibinfo {volume}
  {77}},\ \bibinfo {pages} {114028} (\bibinfo {year}
  {2008}{\natexlab{b}})}\BibitemShut {NoStop}%
\bibitem [{\citenamefont {Du}\ \emph {et~al.}(2013)\citenamefont {Du},
  \citenamefont {Cui}, \citenamefont {Xia},\ and\ \citenamefont
  {Zong}}]{PhysRevD.88.114019}%
  \BibitemOpen
  \bibfield  {author} {\bibinfo {author} {\bibfnamefont {Y.-L.}\ \bibnamefont
  {Du}}, \bibinfo {author} {\bibfnamefont {Z.-F.}\ \bibnamefont {Cui}},
  \bibinfo {author} {\bibfnamefont {Y.-H.}\ \bibnamefont {Xia}}, \ and\
  \bibinfo {author} {\bibfnamefont {H.-S.}\ \bibnamefont {Zong}},\ }\href
  {\doibase 10.1103/PhysRevD.88.114019} {\bibfield  {journal} {\bibinfo
  {journal} {Phys. Rev. D}\ }\textbf {\bibinfo {volume} {88}},\ \bibinfo
  {pages} {114019} (\bibinfo {year} {2013})}\BibitemShut {NoStop}%
\bibitem [{\citenamefont {Lu}\ \emph {et~al.}(2015)\citenamefont {Lu},
  \citenamefont {Du}, \citenamefont {Cui},\ and\ \citenamefont
  {Zong}}]{Lu2015}%
  \BibitemOpen
  \bibfield  {author} {\bibinfo {author} {\bibfnamefont {Y.}~\bibnamefont
  {Lu}}, \bibinfo {author} {\bibfnamefont {Y.-L.}\ \bibnamefont {Du}}, \bibinfo
  {author} {\bibfnamefont {Z.-F.}\ \bibnamefont {Cui}}, \ and\ \bibinfo
  {author} {\bibfnamefont {H.-S.}\ \bibnamefont {Zong}},\ }\href {\doibase
  10.1140/epjc/s10052-015-3720-2} {\bibfield  {journal} {\bibinfo  {journal}
  {Eur. Phys. J. C}\ }\textbf {\bibinfo {volume} {75}},\ \bibinfo {pages} {495}
  (\bibinfo {year} {2015})}\BibitemShut {NoStop}%
\bibitem [{\citenamefont {Cui}\ \emph {et~al.}(2015)\citenamefont {Cui},
  \citenamefont {Hou}, \citenamefont {Shi}, \citenamefont {Wang},\ and\
  \citenamefont {Zong}}]{CUI2015172}%
  \BibitemOpen
  \bibfield  {author} {\bibinfo {author} {\bibfnamefont {Z.-F.}\ \bibnamefont
  {Cui}}, \bibinfo {author} {\bibfnamefont {F.-Y.}\ \bibnamefont {Hou}},
  \bibinfo {author} {\bibfnamefont {Y.-M.}\ \bibnamefont {Shi}}, \bibinfo
  {author} {\bibfnamefont {Y.-L.}\ \bibnamefont {Wang}}, \ and\ \bibinfo
  {author} {\bibfnamefont {H.-S.}\ \bibnamefont {Zong}},\ }\href {\doibase
  https://doi.org/10.1016/j.aop.2015.03.025} {\bibfield  {journal} {\bibinfo
  {journal} {Ann. Phys.}\ }\textbf {\bibinfo {volume} {358}},\ \bibinfo {pages}
  {172 } (\bibinfo {year} {2015})},\ \bibinfo {note} {school of Physics at
  Nanjing University}\BibitemShut {NoStop}%
\bibitem [{\citenamefont {Mark\'o}\ \emph {et~al.}(2015)\citenamefont
  {Mark\'o}, \citenamefont {Reinosa},\ and\ \citenamefont
  {Sz\'ep}}]{PhysRevD.92.125035}%
  \BibitemOpen
  \bibfield  {author} {\bibinfo {author} {\bibfnamefont {G.}~\bibnamefont
  {Mark\'o}}, \bibinfo {author} {\bibfnamefont {U.}~\bibnamefont {Reinosa}}, \
  and\ \bibinfo {author} {\bibfnamefont {Z.}~\bibnamefont {Sz\'ep}},\ }\href
  {\doibase 10.1103/PhysRevD.92.125035} {\bibfield  {journal} {\bibinfo
  {journal} {Phys. Rev. D}\ }\textbf {\bibinfo {volume} {92}},\ \bibinfo
  {pages} {125035} (\bibinfo {year} {2015})}\BibitemShut {NoStop}%
\bibitem [{\citenamefont {Tanabashi}\ and\ \citenamefont
  {et~al.}(2018)}]{PhysRevD.98.030001}%
  \BibitemOpen
  \bibfield  {author} {\bibinfo {author} {\bibfnamefont {M.}~\bibnamefont
  {Tanabashi}}\ and\ \bibinfo {author} {\bibnamefont {et~al.}} (\bibinfo
  {collaboration} {Particle Data Group}),\ }\href {\doibase
  10.1103/PhysRevD.98.030001} {\bibfield  {journal} {\bibinfo  {journal} {Phys.
  Rev. D}\ }\textbf {\bibinfo {volume} {98}},\ \bibinfo {pages} {030001}
  (\bibinfo {year} {2018})}\BibitemShut {NoStop}%
\bibitem [{\citenamefont {Brodsky}\ \emph {et~al.}(2010)\citenamefont
  {Brodsky}, \citenamefont {Roberts}, \citenamefont {Shrock},\ and\
  \citenamefont {Tandy}}]{PhysRevC.82.022201}%
  \BibitemOpen
  \bibfield  {author} {\bibinfo {author} {\bibfnamefont {S.~J.}\ \bibnamefont
  {Brodsky}}, \bibinfo {author} {\bibfnamefont {C.~D.}\ \bibnamefont
  {Roberts}}, \bibinfo {author} {\bibfnamefont {R.}~\bibnamefont {Shrock}}, \
  and\ \bibinfo {author} {\bibfnamefont {P.~C.}\ \bibnamefont {Tandy}},\ }\href
  {\doibase 10.1103/PhysRevC.82.022201} {\bibfield  {journal} {\bibinfo
  {journal} {Phys. Rev. C}\ }\textbf {\bibinfo {volume} {82}},\ \bibinfo
  {pages} {022201} (\bibinfo {year} {2010})}\BibitemShut {NoStop}%
\bibitem [{\citenamefont {Brodsky}\ \emph {et~al.}(2012)\citenamefont
  {Brodsky}, \citenamefont {Roberts}, \citenamefont {Shrock},\ and\
  \citenamefont {Tandy}}]{PhysRevC.85.065202}%
  \BibitemOpen
  \bibfield  {author} {\bibinfo {author} {\bibfnamefont {S.~J.}\ \bibnamefont
  {Brodsky}}, \bibinfo {author} {\bibfnamefont {C.~D.}\ \bibnamefont
  {Roberts}}, \bibinfo {author} {\bibfnamefont {R.}~\bibnamefont {Shrock}}, \
  and\ \bibinfo {author} {\bibfnamefont {P.~C.}\ \bibnamefont {Tandy}},\ }\href
  {\doibase 10.1103/PhysRevC.85.065202} {\bibfield  {journal} {\bibinfo
  {journal} {Phys. Rev. C}\ }\textbf {\bibinfo {volume} {85}},\ \bibinfo
  {pages} {065202} (\bibinfo {year} {2012})}\BibitemShut {NoStop}%
\end{thebibliography}%
\end{document}